\documentclass[journal, 12pt, onecolumn, draftclsnofoot]{IEEEtran}
\normalsize

\usepackage{cite}
\usepackage[dvips]{graphicx}
\ifCLASSOPTIONcompsoc
\usepackage[caption=false,font=normalsize,labelfon
t=sf,textfont=sf]{subfig}
\else
\usepackage[caption=false,font=footnotesize]{subfi
g}
\fi
\usepackage[cmex10]{amsmath}
\usepackage{amsthm,amssymb}
\usepackage{bigints}
\usepackage{multirow}

\newtheorem{theorem}{Theorem}
\newtheorem{lemma}{Lemma}
\newtheorem{remark}{Remark}
\newtheorem{corollary}{Corollary}

\DeclareMathOperator*{\argmax}{arg\,max}

\begin{document}
%
\title{Mode Selection and Spectrum Partition for D2D Inband Communications: A Physical\\ Layer Security Perspective}
%
%
%
%

\author{Yuanyu Zhang, \IEEEmembership{Member, IEEE},
        Yulong Shen, \IEEEmembership{Member, IEEE},
				Xiaohong Jiang, \IEEEmembership{Senior Member, IEEE}
				and Shoji Kasahara, \IEEEmembership{Member, IEEE}
\thanks{Y. Zhang and S. Kasahara are with the Graduate School of Information Science,
       Nara Institute of Science and Technology, 8916-5 Takayama, Ikoma, Nara, 630-0192, Japan. Email: \{yyzhang,kasahara\}@is.naist.jp. }
\thanks{Y. Shen is with the School of Computer Science and Technology, Xidian University, Shaanxi, China. E-mail:ylshen@mail.xidian.edu.cn.}
\thanks{X. Jiang is with the School of Systems Information Science, Future University Hakodate, Hakodate, Hokkaido, Japan, and the School of Computer Science and Technology, Xidian University, Shaanxi, China. E-mail: jiang@fun.ac.jp.}
}

\maketitle

\begin{abstract}
This paper investigates from the physical layer security (PLS) perspective the fundamental issues of mode selection and spectrum partition in cellular networks with inband device-to-device (D2D) communication. We consider a mode selection scheme allowing each D2D pair to probabilistically switch between the underlay and overlay modes, and also a spectrum partition scheme where the system spectrum is orthogonally partitioned between cellular and overlay D2D communications. We first develop a general theoretical framework to model both the secrecy outage/secrecy capacity performance of cellular users and outage/capacity performance of D2D pairs, and to conduct performance optimization to identify the optimal mode selection and spectrum partition for secrecy capacity maximization and secrecy outage probability minimization. A case study is then provided to demonstrate the application of our theoretical framework for performance modeling and optimization, and also to  illustrate the impacts of mode selection and spectrum partition on the PLS performances of inband D2D communications. 
\end{abstract}
\begin{IEEEkeywords}
Device-to-device, physical layer security, mode selection, spectrum partition.
\end{IEEEkeywords}


%
\IEEEpeerreviewmaketitle

\section {Introduction}\label{sec_intro} 
Device-to-device (D2D) communication, which enables nearby users to communicate directly without traversing the base station, has been recognized as one of the key technologies in 5G cellular communications \cite{Boccardi2014}. D2D communication can bring many benefits like increased spectrum efficiency, enlarged cellular coverage and reduced transmission latency, and also open up new opportunities for various proximity-based services such as social networking, content sharing, multi-player gaming, etc. \cite{asadi2014survey, doppler2009d2d, fodor2012design}.

Depending on the spectrum used by D2D pairs, the D2D communication can be classified into inband D2D on cellular spectrum and outband D2D on unlicensed spectrum (e.g., Bluetooth \cite{jo2015device} and WiFi Direct \cite{camps2013device}). This paper focuses on the inband D2D, which improves the overall spectrum efficiency compared to the outband one \cite{doppler2009d2d, fodor2012design}. In the inband D2D, a potential D2D pair can operate either in cellular mode relaying their messages through the base station or in D2D mode communicating directly. The D2D mode can be further divided into overlay D2D using dedicated spectrum resources and underlay D2D reusing the spectrum resources of cellular users (CUEs). As a result, two fundamental and interrelated issues arise in cellular networks with inband D2D communication. The first one is mode selection, i.e., the process of determining which mode each D2D pair should operate in, and the second one is spectrum partition, i.e., how to partition the system spectrum between cellular and overlay D2D communications. This paper jointly considers these two issues in cellular networks with inband D2D communication. In particular, we aim to investigate the optimal mode selection and spectrum partition there from the physical layer security perspective. 

Physical layer security (PLS), which exploits the inherent randomness of wireless channels and noise to provide a strong form of security guarantee, has been identified as a highly promising security approach for 5G cellular communications \cite{Mukherjee2014, NYang2015}. One typical PLS technique is cooperative jamming, which utilizes the artificial noise from helping nodes (also known as friendly jammers) to create a relatively better legitimate channel than the eavesdropping channel \cite{Bassily2013SPM,zhang2015AHWSN, Zhang2015TSC,Zhang2016TASE}. In cellular networks with inband D2D communication, underlay D2D pairs are allowed to reuse the spectrum resource of CUEs, causing interference to CUEs. Such interference is conventionally regarded as an obstacle that degrades the cellular performances and  hinders the application of D2D communication in 5G cellular systems \cite{Tehrani2014IEEEComMag, ChMaTCom2016, Alihemmati2017TWC}. However, from the PLS perspective, the interference from underlay D2D pairs can play the similar role as the artificial noise in cooperative jamming to protect cellular communications from being intercepted by eavesdroppers. 

Motivated by this observation, extensive research efforts have been devoted to the study of inband D2D communication from the PLS perspective, such as  security-oriented resource sharing between underlay D2D pairs and CUEs \cite{Yue2013,JWang2016,HZhang2014ICC,RZhang2016TWC,LWang2015GLOBECOM,Zhang2017TVT},  PLS performance evaluation of D2D underlaying cellular networks \cite{ChMa2015, YLiu2016TOC,Tolossa2017},  security-constrained interference management \cite{Sun2016TvT}, etc. These works demonstrate the potentials of inband D2D communication in enhancing the PLS performances of cellular networks, but the fundamental mode selection and spectrum partition issues were largely ignored therein. Actually, these two issues have significant impacts on the system PLS performances. For example, different settings of mode selection lead to different densities of underlay D2D pairs (i.e., jammers) and different settings of spectrum partition result in different amount of spectrum available to CUEs and overlay D2D pairs, which greatly affects the PLS performances of CUEs and  D2D pairs. However, these impacts remain largely unexplored due to the lack of a joint study on the mode selection and spectrum partition issues from the PLS perspective. Although there have been joint studies on these two issues in \cite{ye2014twc,lin2014TWC,Afzal2016WCNC,Chun2017TWC,zhu2015TWC,Huang2016TCOM,Feng2015TWC} with the objective of maximizing the system overall rate or energy efficiency (Please refer to Section \ref{sec_related_work} for related works), the security issue was not considered therein. Consequently, their results cannot be readily applied to model and optimize the system PLS performances. Therefore, a new and dedicated research is deserved to investigate the impacts of mode selection and spectrum partition on the PLS performances of cellular networks with inband D2D communication, and to further determine the optimal mode selection and spectrum partition settings from the PLS perspective. 

To address this issue, this paper develops a general theoretical framework to model both the PLS performances of CUEs and reliable communication performances of D2D pairs, and to identify the optimal mode selection and spectrum partition for performance optimization. To the best of our knowledge, this is the first work that jointly studies the mode selection and spectrum partition issues from the PLS perspective. The main contributions are summarized as follows.

\begin{itemize} 
\item This paper considers a cellular network with one base station, multiple D2D pairs and multiple CUEs whose transmissions are overheard by an eavesdropper. We adopt the probabilistic mode selection scheme, where each D2D pair independently selects with certain probability to operate in either the underlay mode or overlay mode, and the orthogonal spectrum partition scheme, where the system spectrum is orthogonally divided into two fractions and each fraction is equally shared among the CUEs and among the overlay D2D pairs, respectively. The underlay D2D pairs reuse the spectrum of the CUEs and simultaneously act as friendly jammers to protect the cellular communications. For this network scenario, we derive the theoretical models for the secrecy outage probability (SOP) and average secrecy capacity (ASC) of the CUEs as well as the outage probability (OP) and average capacity (AC) of the D2D pairs. 

\item With the help of above theoretical models, we conduct performance optimization to identify the optimal mode selection probabilities and spectrum partition factors to maximize a weighted proportional fair function in terms of the sum ASC of CUEs and the sum AC of D2D pairs, and to minimize a weighted proportional fair function in terms of the sum SOP of CUEs and the sum OP of D2D pairs, respectively.

\item A case study under the scenario with one CUE and one D2D pair is provided to demonstrate the application of our theoretical framework for performance modeling and optimization, and numerical results for the case study are further presented to illustrate the impacts of mode selection probability and spectrum partition factor on the system performances.
\end{itemize}

The remainder of the paper is organized as follows. Section  \ref{sec_related_work} presents the related works and  Section \ref{sec_model} introduces the system model, mode selection/spectrum partition scheme, and performance metrics in this paper. Section \ref{sec_analysis} presents our theoretical framework for performance modeling and optimization. Section \ref{sec_case_study} provides the case study to illustrate the application of our theoretical framework. Numerical results and the corresponding discussions are presented in Section \ref{sec_num_dis}. Finally, Section \ref{sec_con} concludes this paper.

\section{Related Work}\label{sec_related_work} 
For the inband D2D-enabled cellular networks, available works on the joint study of mode selection and spectrum partition issues without the consideration of security can be roughly classified into two categories. In the first category, mode selection is performed between cellular and overlay D2D modes, while in the second category mode selection is performed among cellular, overlay D2D and underlay D2D modes simultaneously. 

Regarding the works in the first category, Ye \emph{et al.} \cite{ye2014twc} investigated the optimal \emph{probabilistic mode selection} and \emph{orthogonal spectrum partition} to jointly maximize the total rate of potential D2D pairs and CUEs per unit area in a cellular network, where CUEs, potential D2D pairs and base stations are distributed according to Poisson Point Processes (PPPs). Based on the same orthogonal spectrum partition scheme in \cite{ye2014twc} and a new \emph{D2D distance-based mode selection} scheme, Lin \emph{et al.} \cite{lin2014TWC} studied the optimal D2D distance threshold and spectrum partition factor to maximize a weighted proportional fair function in terms of the average rates of potential D2D pairs and CUEs. Later, the authors in \cite{Afzal2016WCNC} extended \cite{lin2014TWC} by considering a more practical path loss model and studied the cellular and D2D coverage probabilities as well as the average network throughput. The authors in \cite{Chun2017TWC} also extended \cite{lin2014TWC} by adopting two generalized fading models and evaluated the spectrum efficiency and outage probability performances of potential D2D pairs and CUEs, respectively. 

Regarding the works in the second category, Zhu \emph{et al.} \cite{zhu2015TWC} proposed a dynamic Stackelberg game framework to identify the optimal mode selection and orthogonal spectrum partition in a finite cellular network, where potential D2D pairs and CUEs are distributed according to PPPs. For a two-tier cellular network with a potential D2D pair, a macro base station and a femto access point, the authors in \cite{Huang2016TCOM} proposed a mode selection scheme based on the D2D distance, received interference of the D2D pair and the availability of orthogonal spectrum resource, and further explored the optimal spectrum partition issues under both the overlay and cellular D2D modes. Different from \cite{ye2014twc,lin2014TWC,Afzal2016WCNC,Chun2017TWC,zhu2015TWC,Huang2016TCOM}, which explored the mode selection and spectrum partition issues from the perspective of sum rate maximization, the authors in \cite{Feng2015TWC} investigated the optimal mode selection and spectrum partition to maximize the overall energy efficiency in a network with one base station, one CUE and one potential D2D pair. It is notable that this paper differs from the above works by jointly investigating the mode selection and spectrum partition issues from the PLS perspective.

\section{Preliminaries} \label{sec_model} 
\subsection{System Model}
\begin{figure}[!t]
\centering
\includegraphics[width=3in]{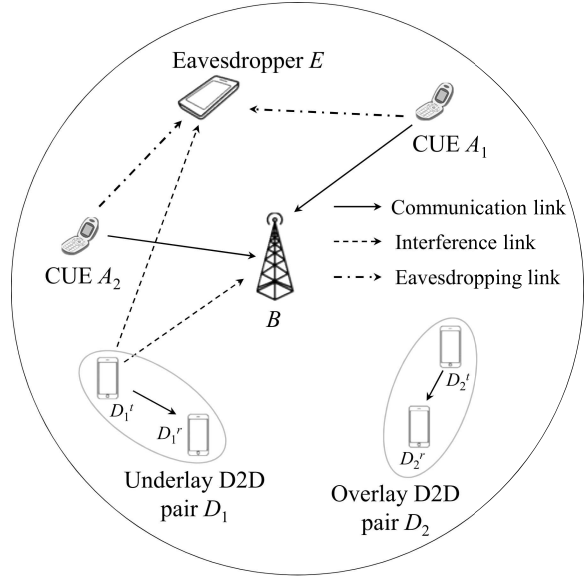}
\caption{System model for an isolated cell with inband D2D pairs.}
\label{fig_sysmodel}
\end{figure}

As illustrated in Fig. \ref{fig_sysmodel}, we consider a cellular network consisting of one base station $B$, one eavesdropper $E$, $n$ cellular users (CUEs) $\mathcal A=\{A_1,A_2,\cdots,A_n\}$ and $m$ D2D pairs $\mathcal D=\{D_1,D_2,\cdots,D_m\}$. We focus on the uplink transmissions of CUEs\footnote{Although we focus on the uplink scenario in this paper, our results also apply to the downlink scenario.}, as sharing the uplink resource with D2D pairs offers several benefits like improved spectrum utilization and better interference management \cite{Lin2014CoMMag}. We use $D_j^{\,t}$ and $D_j^{\,r}$ to denote the transmitter and receiver of the $j$-th D2D pair, respectively. We assume that the eavesdropper $E$ overhears the transmissions of all CUEs and only its statistical channel state information (CSI) is known. Each node has a single omnidirectional antenna, and the CUE $A_i$ ($i\in\{1,2,\cdots,n\}$) and D2D transmitter $D_j^{\,t}$ ($j\in{1,2,\cdots, m}$) transmit with power $P_{A_i}$ and $P_{D_j^{\,t}}$, respectively. We assume all CUEs (resp. D2D transmitters) adopt a common transmit power, i.e., $P_{A_i}=P_A$ (resp. $P_{D_j^{\,t}}=P_D$). We consider a time-slotted system and a quasi-static Rayleigh fading channel model where each channel remains static for one slot but changes randomly and independently from slot to slot. The channel coefficient between nodes $i$ and $j$ is denoted as $h_{i,j}$, which is modeled as a complex zero mean Gaussian random variable with variance $\sigma^2_{i,j}=d_{i,j}^{-\alpha}$, where $\alpha$ is the path-loss exponent and $d_{i,j}$ is the Euclidean distance between $i$ and $j$. Thus, the corresponding channel gain $|h_{i,j}|^2$ is an exponentially distributed random variable with mean $d_{i,j}^{-\alpha}$. In addition, all wireless channels are impaired by additive white Gaussian noise with variance $\sigma^2$. We assume that the available system spectrum has a total bandwidth of $W$ MHz. Without loss of generality, we assume $W=1$ throughout this paper.  

\subsection{Mode Selection and Spectrum Partition}\label{sec_mssp}
This paper considers two communication modes for D2D pairs, i.e., \textbf{\emph{overlay mode}} where D2D pairs use dedicated spectrum resource, and \textbf{\emph{underlay mode}} where D2D pairs reuse the spectrum resource of CUEs. We consider a probabilistic mode selection scheme where each D2D pair independently and randomly selects to operate in the underlay mode with probability $p$ (and thus in the overlay mode with probability $1-p$) in each time slot. We use $\mathcal D_u$ and $\mathcal D_o$ to represent the set of D2D pairs operating in the underlay mode and in the overlay mode, respectively. We adopt the orthogonal spectrum partition scheme, which partitions the system spectrum into two fractions: a fraction $\beta$ of the spectrum is orthogonally and equally shared among the CUEs and the remaining fraction is orthogonally and equally shared among the overlay D2D pairs in $\mathcal D_o$\footnote{$\beta=1$ if no D2D pair operates in the overlay mode (i.e., $\mathcal D_o=\emptyset$); otherwise, $\beta \in [0,1]$.}. The underlay D2D pairs in $\mathcal D_u$ reuse the spectrum resource allocated to the CUEs and simultaneously act as friendly jammers to protect the CUEs from the eavesdropping attack of the eavesdropper $E$. We assume that each underlay D2D pair in $\mathcal D_u$ is allowed to independently and randomly reuse the resource of \emph{only one} CUE with equal probability $1/n$.

\subsection{Performance Metrics}
Consider the uplink transmission of the $i$-th CUE $A_i$, the instantaneous signal-to-interference-plus-noise ratio (SINR) at the base station $B$ is given by
\begin{IEEEeqnarray}{rCl}  
\mathrm{SINR}_{A_i,B}=\frac{P_{A_i}|h_{A_i,B}|^2}{\sum_{D_k\in \mathcal D_u^i} P_{D_{k}^{\,t}}|h_{D_{k}^{\,t},B}|^2+\sigma^2}=\frac{\mathrm{SNR}_{A_i,B}}{\sum_{D_k\in\mathcal D_u^i}\mathrm{SNR}_{D_{k}^{\,t},B}+1},
\end{IEEEeqnarray}
where $\mathcal D_u^i\subseteq\mathcal D_u$ denotes the set of D2D pairs reusing the spectrum of CUE $A_i$, $\mathrm{SNR}_{a,b}=P_a|h_{a,b}|^2/\sigma^2$ ($a\in\{A_i, D_{k}^{\,t}\}$ and $b\in\{B\}$) denotes the signal-to-noise ratio (SNR) from nodes $a$ to $b$. Note that $\mathrm{SNR}_{a,b}$ is an exponentially distributed random variable with mean $\gamma_{a,b}=P_ad_{a,b}^{-\alpha}/\sigma^2$. Similarly, the instantaneous SINR at the eavesdropper $E$ is given by  
\begin{align} 
\mathrm{SINR}_{A_i,E}=\frac{\mathrm{SNR}_{A_i,E}}{\sum_{D_k\in\mathcal D_u^i}\mathrm{SNR}_{D_{k}^{\,t},E}+1}.
\end{align}
According to \cite{Mukherjee2014,Bassily2013SPM}, the instantaneous secrecy capacity $C_s^{\,i}$ of CUE $A_i$ is determined as
\begin{align}
C_s^{\,i} =\frac{\beta}{n}\left[\log\left(\frac{1+\mathrm{SINR}_{A_i,B}}{1+\mathrm{SINR}_{A_i,E}}\right)\right]^+,
\end{align}
where $[x]^+=\max\{x,0\}$ and $\log$ is to the base of $2$. 
 
To model the security performance of CUE $A_i$, we adopt the metric of \textbf{\emph{secrecy outage probability}} (SOP) to characterize the probability that the instantaneous secrecy capacity of $A_i$ falls below a target secrecy rate $r_s$. Formally, we formulate the SOP $\mathbf P_{so}^{\,i}$ of CUE $A_i$ as
\begin{align}\label{sop_def}
\mathbf P_{so}^{\,i} = \mathbb P\left(C_s^{\,i} < r_s\right),
\end{align}
where $\mathbb P(\cdot)$ represents the probability operator. We also adopt the metric of \textbf{\emph{average secrecy capacity}} (ASC) to depict the expected maximum achievable secrecy rate of $A_i$, which is denoted as $\mathbf C_s^{\,i}$ and given by 
\begin{align}\label{sc_def}
\mathbf C_s^{\,i} = \mathbb E[C_s^{\,i}],
\end{align} 
where $\mathbb E[\cdot]$ represents  the expectation operator.

For the $j$-th D2D pair $D_j$, when it operates in the underlay mode and reuses the spectrum of CUE $A_i$, the instantaneous SINR from the transmitter $D_{j}^{t}$ to the receiver $D_j^{r}$ is given by
\begin{align}
\mathrm{SINR}_{D_{j}^{t},D_j^{r}}\!=\!\frac{\mathrm{SNR}_{D_{j}^{t},D_j^{r}}}{\mathrm{SNR}_{A_i,D_{j}^{r}}\!+\!\underset{D_k\in \mathcal D_u^i\backslash \{D_j\}}{\sum}\mathrm{SNR}_{D_{k}^{\,t},D_{j}^{r}}\!+\!1}.
\end{align}
Thus, the instantaneous capacity $R_{u}^{j,\,i}$ of  $D_j$ in the underlay mode is given by 
\begin{align}
R_{u}^{j,\,i}=\frac{\beta}{n}\log\left(1+\mathrm{SINR}_{D_{j}^{t},D_{j}^{r}}\right).
\end{align}
When $D_j$ operates in the overlay mode, it orthogonally and equally shares the allocated spectrum with other overlay D2D pairs in $\mathcal D_o$, so no interference from other D2D pairs will exist during its transmission. The instantaneous capacity $R_o^{\,j}$ of $D_j$ in the overlay mode is given by
\begin{align}
R_o^{\,j}=\frac{(1-\beta)}{|\mathcal D_o|}\log\left(1+\mathrm{SNR}_{D_{j}^{t},D_j^{r}}\right).
\end{align}
  
To model the performance of D2D pair $D_j$, we adopt the metric of \textbf{\emph{outage probability}} (OP), which is defined as the probability that the instantaneous capacity of $D_j$ falls below a target rate $r_t$. Formally, we formulate the OP $\mathbf P_o^{j}$ of D2D pair $D_j$ as
\begin{align}\label{op_def}
\mathbf P_o^{j}=\frac{p}{n}\sum_{i=1}^{n}\mathbb P(R_{u}^{j,\,i}<r_t)+(1-p)\mathbb P(R_o^{\,j}<r_t),
\end{align} 
where $\mathbb P(R_{u}^{j,\,i}<r_t)$ denotes the OP when $D_j$ operates in the underlay mode and reuses the resource of CUE $A_i$, and $\mathbb P(R_o^{\,j}<r_t)$ denotes the OP when $D_j$ operates in the overlay mode. To characterize the expected maximum achievable rate of $D_j$, we adopt the metric of \textbf{\emph{average capacity}} (AC), which is denoted as $\mathbf R_j$ and given by
\begin{align}\label{ac_def}
\mathbf R_j = \frac{p}{n}\sum_{i=1}^{n}\mathbb E[R_{u}^{j,\,i}]+(1-p)\mathbb E[R_o^{\,j}],
\end{align}
where $\mathbb E[R_{u}^{j,\,i}]$ denotes the AC when $D_j$ operates in the underlay mode and reuses the resource of CUE $A_i$, and $\mathbb E[R_o^{\,j}]$ denotes the AC when $D_j$ operates in the overlay mode.  
 
\section{Performance Modeling and Optimization} \label{sec_analysis} 
This section presents our general theoretical framework for the performance modeling and optimization. In this framework, we first derive the theoretical models for the SOP and ASC of CUEs as well as the OP and AC of D2D pairs, based on which we then study the optimal settings of mode selection probability and spectrum partition factor for performance optimization. 
\subsection{SOP and ASC of CUEs}
For a CUE $A_i$ ($i\in\{1,2,\cdots, n\}$), we define $I_{B}=\sum_{D_k\in\mathcal D_u^i}\mathrm{SNR}_{D_{k}^{\,t},B} $ and $I_E=\sum_{D_k\in\mathcal D_u^i}\mathrm{SNR}_{D_{k}^{\,t},E}$ as the total interference at the base station and at the eavesdropper $E$, respectively. We can see that both $I_B$ and $I_E$ are the sums of a random number of independent random variables and their pdfs vary with different realizations of $\mathcal D_u^i$. We use $\mathbf D_u$ to denote a particular realization of $\mathcal D_u^i$. It is easy to see that $\mathbf D_u\in 2^{\mathcal D}$, where $2^{\mathcal D}$ denotes the power set of $\mathcal D$. 
 
Before giving the main results, we first introduce two basic pdfs $\mathbf f_{\mathbf D_u}^i(x)$ and $\mathbf g_{\mathbf D_u}^i(x)$ regarding the $I_B$ and $I_E$ for a particular $\mathbf D_u$, respectively. For a given $\mathbf D_u$, $I_B$ and $I_E$ are the sum of multiple independent random variables. In general, the analytical expressions of $\mathbf f_{\mathbf D_u}^i(x)$ and $\mathbf g_{\mathbf D_u}^i(x)$ are difficult to obtain, but they can be determined by the following multi-fold convolutions
\begin{align}\label{pdf_fx}
\mathbf f_{\mathbf D_u}^i(x)= \underbrace{(f_{k}*\cdots*f_{l})}_{|\mathbf D_u|}(x),\,\mathbf g_{\mathbf D_u}^i(x)= \underbrace{(g_{k}*\cdots*g_{l})}_{|\mathbf D_u|}(x),
\end{align}
where $f_k$ and $f_l$ are the pdfs of  $\mathrm{SNR}_{D_{k}^{\,t},B}$ and $\mathrm{SNR}_{D_{l}^{\,t},B}$, and $g_k$ and $g_l$ are the pdfs of $\mathrm{SNR}_{D_{k}^{\,t},E}$ and $\mathrm{SNR}_{D_{l}^{\,t},E}$ for $D_k, D_l \in\mathbf D_u$. For the special case where all $\mathrm{SNR}_{D_{k}^{\,t},B}$ (resp. $\mathrm{SNR}_{D_{k}^{\,t},E}$) are independently and identically distributed with parameter $\lambda$, $\mathbf f_{\mathbf D_u}^i(x)$ (resp. $\mathbf g_{\mathbf D_u}^i(x)$) can be analytically given by the pdf of an Erlang distribution with parameters $|\mathbf D_u|$ and $\lambda$. Based on $\mathbf f_{\mathbf D_u}^i(x)$ and $\mathbf g_{\mathbf D_u}^i(x)$, we are now ready to give the main results for the SOP and ASC of CUE $A_i$ in the following theorem. 

\begin{theorem} \label{Theorem_CUE} Consider the network scenario as shown in Fig. \ref{fig_sysmodel} with one base station, $n$ CUEs, $m$ D2D pairs and one eavesdropper, the SOP of CUE $A_i$ ($i\in\{1,2,\cdots, n\}$) under the mode selection and spectrum partition schemes introduced in Section \ref{sec_mssp} is given by
\begin{align}\label{theorem_sop}
\mathbf P_{so}^{\,i}=1-\sum_{\mathbf D_u\in 2^{\mathcal D}}\varepsilon^{|\mathbf D_u|}\,\left[\vartheta^{m-|\mathbf D_u|}\Theta_{i}(\mathbf D_u,1)+\left(\left(1\!-\!\varepsilon\right)^{m-|\mathbf D_u|}\!-\!\vartheta^{m-|\mathbf D_u|}\right)\Theta_{i}(\mathbf D_u,\beta)\right],
\end{align}
where $\varepsilon= \frac{p}{n}$ denotes the probability that an underlay D2D pair reuses the resource of CUE $A_i$, $\vartheta=p\left(1-\frac{1}{n}\right)$ denotes the probability that an underlay D2D pair reuses the resource of other CUEs except for $A_i$ and 
\begin{align}\label{eqn_theta_empty}
\Theta_{i}(\mathbf D_u,\beta)=\frac{e^{-\frac{2^{\frac{nr_s}{\beta}}-1 }{\gamma_{A_i,B}}}}{\frac{\gamma_{A_i,E}}{\gamma_{A_i,B}}2^{\frac{nr_s}{\beta }}+1},
\end{align} 
if $\mathbf D_u=\emptyset$; otherwise,  
\begin{align}\label{eqn_theta_nonempty}
\Theta_{i}(\mathbf D_u,\beta)=\int_{0}^{\infty}\!\!\!\int_{0}^{\infty}\!\!\! \frac{e^{-\frac{(2^{\frac{nr_s}{\beta }}-1) (x+1)}{\gamma_{A_i,B}}}}{\frac{\gamma_{A_i,E}}{\gamma_{A_i,B}}\frac{x+1}{y+1}2^{\frac{nr_s}{\beta}}+1}\mathbf f_{\mathbf D_u}^i(x)\mathbf g_{\mathbf D_u}^i(y)\mathrm{d}x\mathrm{d}y.
\end{align}
The ASC of CUE $A_i$ is given by 
\begin{IEEEeqnarray}{rCl}\label{theorem_asc}
\mathbf C_s^{\,i} = \frac{1}{n \ln2 }\sum_{\mathbf D_u\in 2^{\mathcal D}}\!\varepsilon^{|\mathbf D_u|}\Lambda_{i}(\mathbf D_u)\left(\beta\left(1\!-\!\varepsilon\right)^{m-|\mathbf D_u|}+(1\!-\!\beta)\vartheta^{m-|\mathbf D_u|}\right),
\end{IEEEeqnarray} 
where 
\begin{IEEEeqnarray}{rCl}\label{eqn_lambda_empty}
\Lambda_{i}(\mathbf D_u)=\Psi\left(\frac{1}{\gamma_{A_i,B}}+\frac{1}{\gamma_{A_i,E}}\right)-\Psi\left(\frac{1}{\gamma_{A_i,B}}\right),
\end{IEEEeqnarray} 
if $\mathbf D_u=\emptyset$; otherwise,
\begin{IEEEeqnarray}{rCl}\label{eqn_lambda_nonempty}
\Lambda_{i}(\mathbf D_u)&=&\int_0^{\infty}\!\!\!\!\int_{0}^{\infty} \Bigg[\Psi\left(\frac{x+1}{\gamma_{A_i,B}}+\frac{y+1}{\gamma_{A_i,E}}\right)-\Psi\left(\frac{x+1}{\gamma_{A_i,B}}\right)\Bigg]\mathbf f_{\mathbf D_u}^i(x)\mathbf g_{\mathbf D_u}^i(y)\mathrm{d}x\mathrm{d}y,
\end{IEEEeqnarray} 
where $\Psi(x)=e^x\mathrm{Ei}(-x)$ and  $\mathrm{Ei}(x)=-\int_{-x}^{\infty}\frac{e^{-t}}{t}\mathrm{d}t$ denotes the exponential integral. 
\end{theorem}
\begin{IEEEproof}
See Appendix \ref{app_theorem_cue}.
\end{IEEEproof}

\subsection{OP and AC of D2D Pairs} 
This subsection provides the general expressions for the OP and AC of D2D pairs. For a D2D pair $D_j$ ($j\in\{1,2,\cdots, m\}$) reusing the spectrum of CUE $A_i$, we define $I_{D_j^{r}}^i=\sum_{D_k\in \mathcal D_u^i\backslash \{D_j\}}\mathrm{SNR}_{D_{k}^{\,t},D_{j}^{r}}$ as the interference at the D2D receiver $D_{j}^{r}$. We use $\mathbf D_u^j $ to denote a particular realization of $\mathcal D_u^i\backslash \{D_j\}$. It is easy to see that $\mathbf D_u^j\in 2^{\mathcal D^{'}}$, where $2^{\mathcal D^{'}}$ is the power set of $\mathcal D^{'}=\mathcal D\backslash\{D_j\}$. We can see that, for a given realization of $\mathbf D_u^j$, $I_{D_j^{r}}^i$ is the sum of $|\mathbf D_u^j|$ independent random variables. We use $\mathbf h_{\mathbf D_u^j}^j(x)$ to denote the pdf of $I_{D_j^{r}}^i$ for a given $\mathbf D_u^j$. Similar to $\mathbf f_{\mathbf D_u}^i(x)$ and $\mathbf g_{\mathbf D_u}^i(x)$, $\mathbf h_{\mathbf D_u^j}^j(x)$ can be determined by the following multi-fold convolution, 
\begin{IEEEeqnarray}{rCl}\label{pdf_hx} 
\mathbf h_{\mathbf D_u^j}^j(x)= \underbrace{(h_{k}*\cdots*h_{l})}_{|\mathbf D_u^j|}(x),  
\end{IEEEeqnarray} 
where $h_k$ and $h_l$ denote the pdfs of $\mathrm{SNR}_{D_{k}^{\,t},D_j^{r}}$ and $\mathrm{SNR}_{D_{l}^{\,t},D_j^{r}}$ for $D_k,D_l \in\mathbf D_u^j$, respectively. Based on $\mathbf h_{\mathbf D_u^j}^j(x)$, we give the following theorem regarding the OP and AC of D2D pair $D_j$.
\begin{theorem} \label{Theorem_D2D}
Consider the network scenario as shown in Fig. \ref{fig_sysmodel} with one base station, $n$ CUEs, $m$ D2D pairs and one eavesdropper, the OP of D2D pair $D_j$ ($j\in\{1,2,\cdots, m\}$) under the mode selection and spectrum partition schemes introduced in Section \ref{sec_mssp} is given by
\begin{IEEEeqnarray}{rCl} \label{theorem_op}
\mathbf P_{o}^j&=&1-\sum_{l=1}^{m}\binom{m-1}{l-1}p^{m-l}(1-p)^{l}e^{-\frac{2^{\frac{lr_t}{(1-\beta)}}-1}{\gamma_{D_{j}^{t},D_j^{r}}}}\\
&&\!-\!\sum_{i=1}^{n}\!\Bigg[\sum_{\mathbf D_u^j\in 2^{\mathcal D^{'}}}\! \varepsilon^{|\mathbf D_u^j|+1}\left[\vartheta^{m-1-|\mathbf D_u^j|}\Omega_{i,j}(\mathbf D_u^j,1)\!+\!\left((1\!-\!\varepsilon)^{m\!-\!1\!-\!|\mathbf D_u^j|}\!-\!\vartheta^{m-1-|\mathbf D_u^j|}\right)\Omega_{i,j}(\mathbf D_u^j,\beta)\right]\Bigg]\nonumber,
\end{IEEEeqnarray} 
where
\begin{IEEEeqnarray}{rCl}\label{eqn_omega_empty}
\Omega_{i,j}(\mathbf D_u^j,\beta)=\frac{e^{-\frac{2^{\frac{nr_t}{\beta }}-1}{\gamma_{D_{j}^{t},D_j^{r}}}}}{\frac{\gamma_{A_i,D_{j}^{r}}}{\gamma_{D_{j}^{t},D_j^{r}}}\left(2^{\frac{nr_t}{\beta }}-1\right)+1},
\end{IEEEeqnarray}
if $\mathbf D_u^j=\emptyset$; otherwise,
\begin{IEEEeqnarray}{rCl}\label{eqn_omega_non_empty}
\Omega_{i,j}(\mathbf D_u^j,\beta)&=&\frac{\int_{0}^{\infty}e^{-\frac{\left(2^{\frac{nr_t}{\beta }}-1\right)(x+1)}{\gamma_{D_{j}^{t},D_j^{r}}}}\mathbf h_{\mathbf D_u^j}^j(x)\mathrm{d}x}{\frac{\gamma_{A_i,D_{j}^{r}}}{\gamma_{D_{j}^{t},D_j^{r}}}\left(2^{\frac{nr_t}{\beta }}-1\right)+1}.
\end{IEEEeqnarray}
The AC of D2D pair $D_j$ is given by 
\begin{IEEEeqnarray}{rCl}\label{theorem_ac}
\mathbf R_j &=&\sum_{i=1}^{n}\sum_{\mathbf D_u^j\in 2^{\mathcal D^{'}}}\frac{\!\!\varepsilon^{|\mathbf D_u^j|+1}\Delta_{i,j}(\mathbf D_u^j)
\left[\beta(1-\varepsilon)^{m-1-|\mathbf D_u^j|}+(1-\beta)\vartheta^{m-1-|\mathbf D_u^j|}\right]}{n\ln 2\left(\frac{\gamma_{A_i,D_{j}^{r}}}{\gamma_{D_{j}^{t},D_j^{r}}}-1\right)}\\
&& -\frac{(1\!-\!\beta)\Psi\left(\frac{1}{\gamma_{D_j^{t},D_j^{r}}}\right)}{\ln 2}\sum_{l=1}^{m}\binom{m\!-\!1}{l\!-\!1}\frac{p^{m-l}(1\!-\!p)^{l}}{l },\nonumber
\end{IEEEeqnarray}
where 
\begin{IEEEeqnarray}{rCl}\label{eqn_delta_empty}
\Delta_{i,j}(\mathbf D_u^j)=\Psi\left(\frac{1}{\gamma_{D_{j}^{t},D_j^{r}}}\right)-\Psi\left(\frac{1}{\gamma_{A_i,D_{j}^{r}}}\right),
\end{IEEEeqnarray}
if $\mathbf D_u^j=\emptyset$; otherwise,
\begin{IEEEeqnarray}{rCl}\label{eqn_delta_nonempty}
\Delta_{i,j}(\mathbf D_u^j)\!=\!\int_{0}^{\infty}\!\!\!\!\mathbf h_{\mathbf D_u^j}^j(x)\left[\Psi\left(\frac{x+1}{\gamma_{D_{j}^{t},D_j^{r}}}\right)\!-\!\Psi\left(\frac{x+1}{\gamma_{A_i,D_{j}^{r}}}\right)\right]\mathrm{d}x.
\end{IEEEeqnarray}
\end{theorem}
\begin{IEEEproof}
See Appendix \ref{app_theorem_d2d}.
\end{IEEEproof}
\begin{remark}
Notice that $\mathbf f_{\mathbf D_u}^i(x)$, $\mathbf g_{\mathbf D_u}^i(x)$ and $\mathbf h_{\mathbf D_u^j}^j(x)$ serve as three fundamental pdfs in our theoretical framework. Once they are determined, the analytical expressions for $\mathbf P_{so}^{\,i}$, $\mathbf C_s^{\,i}$, $\mathbf P_{o}^j$ and $\mathbf R_j $ can be determined accordingly. 
\end{remark}

\subsection{Performance Optimization}\label{sec_opt_formulation}
Based on the results in Theorems $\ref{Theorem_CUE}$ and $\ref{Theorem_D2D}$, this subsection investigates the optimal settings of mode selection probability and spectrum partition factor for system performance optimization. In general, optimization problems in D2D-enabled cellular networks need to consider the fairness between the performances of CUEs and D2D pairs. To do this, we adopt the following weighted proportional fair function  \cite{lin2014TWC}
\begin{align}
\mathcal U(U_c,U_d) = w_c \ln U_c +w_d \ln U_d, 
\end{align}
where $w_c+w_d=1$, and $U_c$ and $U_d$ represent the utilities of CUEs and D2D pairs, respectively. 

We first explore the optimal settings from the perspective of sum rate maximization, for which we consider the following optimization problem (referred to as Problem P1)
\begin{align}\label{fn_objective}
\mathbf {P1}: (p_1^*,\beta_1^*) = \underset{p, \beta\in[0,1]}{\argmax}\,\, \mathcal U \left(\sum_{i=1}^{n} \mathbf C_s^{\,i}, \sum_{j=1}^{m}\mathbf R_j \right),
\end{align} 
where $p_1^*$ and $\beta_1^*$ denote the optimal values of mode selection probability and spectrum partition factor for Problem P1, respectively. To make this problem more explicit, we rewrite $\mathbf C_s^{\,i}$ as
\begin{align}\label{eqn_cs_r}
\mathbf C_s^{\,i} =\frac{a_i\beta+b_i}{n \ln 2},
\end{align} 
where
\begin{IEEEeqnarray}{rCl}
a_i = \!\sum_{\mathbf D_u\in 2^{\mathcal D}}\!\varepsilon^{|\mathbf D_u|}\Lambda_{i}(\mathbf D_u)\left(\left(1\!-\!\varepsilon\right)^{m-|\mathbf D_u|}\!-\!\vartheta^{m-|\mathbf D_u|}\right), b_i = \sum_{\mathbf D_u\in 2^{\mathcal D}}\!\varepsilon^{|\mathbf D_u|}\Lambda_{i}(\mathbf D_u)\vartheta^{m-|\mathbf D_u|}.
\end{IEEEeqnarray}
Notice that $\varepsilon>0$, $1-\varepsilon>\vartheta>0$ and $\Lambda_{i}(\mathbf D_u)>0$, so we have $a_i>0$ and $b_i>0$.
We also rewrite $\mathbf R_j $ as
\begin{IEEEeqnarray}{rCl}\label{eqn_r_r}
\mathbf R_j &=&\frac{(u_j+s_j)\beta + v_j-s_j}{\ln 2},
\end{IEEEeqnarray} 
where
\begin{IEEEeqnarray}{rCl}
u_j &=& \sum_{i=1}^{n}\sum_{\mathbf D_u^j\in 2^{\mathcal D^{'}}}\frac{\varepsilon^{|\mathbf D_u^j|+1}\Delta_{i,j}(\mathbf D_u^j)\left((1-\varepsilon)^{m-1-|\mathbf D_u^j|}-\vartheta^{m-1-|\mathbf D_u^j|}\right)}{n\left(\frac{\gamma_{A_i,D_{j}^{r}}}{\gamma_{D_{j}^{t},D_j^{r}}}-1\right)}\!\!,\\
v_j &=& \sum_{i=1}^{n}\sum_{\mathbf D_u^j\in 2^{\mathcal D^{'}}}\frac{\!\!\varepsilon^{|\mathbf D_u^j|+1}\Delta_{i,j}(\mathbf D_u^j)\vartheta^{m-1-|\mathbf D_u^j|}}{n\left(\frac{\gamma_{A_i,D_{j}^{r}}}{\gamma_{D_{j}^{t},D_j^{r}}}-1\right)},\\
s_j &=& \Psi\left(\frac{1}{\gamma_{D_j^{t},D_j^{r}}}\right)\sum_{l=1}^{m}\binom{m\!-\!1}{l\!-\!1}\frac{p^{m-l}(1\!-\!p)^{l}}{l }.
\end{IEEEeqnarray}
It is easy to see that $\frac{\Delta_{i,j}(\mathbf D_u^j)}{\left(\frac{\gamma_{A_i,D_{j}^{r}}}{\gamma_{D_{j}^{t},D_j^{r}}}-1\right)}> 0$ and $\Psi\left(\frac{1}{\gamma_{D_j^{t},D_j^{r}}}\right)<0$, so we have $u_j>0$, $v_j>0$ and $s_j<0$. From (\ref{eqn_cs_r}) and (\ref{eqn_r_r}), we can see that both $\mathbf C_s^{\,i}$ and $\mathbf R_j $ are linear functions of $\beta$. Based on this property, we are now ready to give the following lemma regarding the optimal spectrum partition factor $\beta_1^* $ of Problem P1.
\begin{lemma}\label{lemma_OP1}
The optimal spectrum partition factor $\beta_1^* $ for Problem P1 is given by 
\begin{IEEEeqnarray}{rCl}
\beta_1^* =\min\left\{ \max\left\{0,-w_d\frac{\sum_{i=1}^{n}b_i}{\sum_{i=1}^{n}a_i}-w_c\frac{\sum_{j=1}^{m}(v_j-s_j)}{\sum_{j=1}^{m}(u_j+s_j)}\right\}, 1\right\}
\end{IEEEeqnarray}
if $\sum_{j=1}^{m}(u_j+s_j)<0$, otherwise, $\beta_1^*=1$. In particular, $\lim_{p\rightarrow 0}\beta_1^*=w_c$.
\end{lemma}
\begin{IEEEproof}
We first take the derivative of $\mathcal U$ with respect to $\beta$, which is given by  
\begin{align}
\frac{\partial \mathcal U}{\partial \beta}=\frac{w_c}{\beta + \frac{\sum_{i=1}^{n}b_i}{\sum_{i=1}^{n}a_i}}+\frac{w_d}{\beta+\frac{\sum_{j=1}^{m}(v_j-s_j)}{\sum_{j=1}^{m}(u_j+s_j)}}.
\end{align}
Recall that $a_i>0$, $b_i>0$, $u_j>0$, $v_j>0$, $s_j<0$, so we have $\sum_{i=1}^{n}a_i>0$, $\sum_{i=1}^{n}b_i>0$ and $\sum_{j=1}^{m}(v_j-s_j)>0$. We can see that the monotonicity of  $\mathcal U$ depends on the sign of $\sum_{j=1}^{m}(u_j+s_j)$. If $\sum_{j=1}^{m}(u_j+s_j)\geq 0$, we have $\frac{\partial \mathcal U}{\partial \beta}>0$ and $\mathcal U$ is increasing with $\beta$. Thus, the optimal $\beta$ is $\beta_1^*=1$. If $\sum_{j=1}^{m}(u_j+s_j)<0$, we have $\sum_{j=1}^{m}(v_j-s_j)+\sum_{j=1}^{m}(u_j+s_j)=\sum_{j=1}^{m}(v_j+u_j)>0$ and thus $\frac{\sum_{j=1}^{m}(v_j-s_j)}{\sum_{j=1}^{m}(u_j+s_j)}<-1$. So, we have $\beta+\frac{\sum_{j=1}^{m}(v_j-s_j)}{\sum_{j=1}^{m}(u_j+s_j)}<0$. Defining $\beta'$ as the solution of $\frac{\partial \mathcal U}{\partial \beta}=0$, we have  
\begin{align}
\beta'= -w_d\frac{\sum_{i=1}^{n}b_i}{\sum_{i=1}^{n}a_i}-w_c\frac{\sum_{j=1}^{m}(v_j-s_j)}{\sum_{j=1}^{m}(u_j+s_j)}.
\end{align}
We can see that $\frac{\partial \mathcal U}{\partial \beta}<0$ for $\beta>\beta'$ and $\frac{\partial \mathcal U}{\partial \beta}>0$ for $\beta<\beta'$. Thus, the optimal $\beta$ is $\beta_1^* = \min\left\{\max\left\{0,\beta'\right\},1\right\}$. In particular, as $p\rightarrow 0$, we have $a_i\rightarrow 0$, $v_j \rightarrow 0$, $u_j \rightarrow 0$ and thus $\beta_1^*\rightarrow w_c$.
\end{IEEEproof}
From Lemma \ref{lemma_OP1}, we can see that as $p\rightarrow 0$, i.e., no D2D pair chooses the underlay mode, the $\beta_1^*$ converges to $w_c$, which is the weight assigned to the utility of CUEs. It is notable that $\beta_1^*$ is a function of  $p$. Substituting $\beta_1^*$ back into (\ref{fn_objective}) reduces $\mathcal U$ to a function of only $p$. Thus, the optimal $p_1^*$ can be found efficiently, which in return determines the value of $\beta_1^*$. 

Next, we explore the optimal settings of mode selection probability and spectrum partition factor from the perspective of outage probability minimization. For this purpose, we consider the following optimization problem (referred to as Problem P2)
\begin{align}\label{fn_objective}
\mathbf {P2}: (p_2^*, \beta_2^*) = \underset{p, \beta\in[0,1]}{\argmax}\,\, -\mathcal U \left(\sum_{i=1}^{n} \mathbf P_{so}^{\,i}, \sum_{j=1}^{m}\mathbf P_{o}^j\right),
\end{align} 
where $p_2^*$ and $\beta_2^*$ denote the optimal values of mode selection probability and spectrum partition factor for Problem P2, respectively. From the expressions of $\mathbf P_{so}^{\,i}$ and $\mathbf P_{o}^j$, we can see that closed-form solutions for $p_2^*$ and $\beta_2^*$ are usually difficult to obtain, so a two-dimensional search over $(p, \beta)$ can be used to find the $p_2^*$ and $\beta_2^*$.

\section{Case Study}\label{sec_case_study}
In this section, we provide a case study to illustrate the application of our theoretical framework for performance modeling and optimization. We consider a simple system with one CUE $A$ and one D2D pair $D$ (i.e., $n=1$ and $m=1$), as considered in \cite{Feng2015TWC,yu2011twc, Cheng2016JSAC}. We first give analytical expressions for the SOP and ASC of CUE $A$ as well as the OP and AC of D2D pair $D$. Based on the analytical expressions, we then solve the related optimization problems to find the optimal settings of mode selection probability and spectrum partition factor.

\subsection{Performance Modeling}
Based on Theorem \ref{Theorem_CUE}, we first provide the analytical expressions for the SOP and ASC of CUE $A$ in the following corollary.
\begin{corollary} \label{corollary_CUE}
Consider a cellular network with one base station $B$, one eavesdropper $E$, one CUE $A$ and one D2D pair $D$, the SOP $\mathbf P_{so}$ of CUE $A$ under the mode selection and spectrum partition schemes introduced in Section \ref{sec_mssp} is given by 
\begin{IEEEeqnarray}{rCl}\label{eqn_case_sop}
\mathbf P_{so}=1-(1-p)\Theta(\emptyset,\beta)-p\,\Theta(D,1),
\end{IEEEeqnarray} 
where $\Theta(\emptyset,\beta)=\frac{e^{-\frac{2^{\frac{r_s}{\beta}}-1 }{\gamma_{A,B}}}}{\frac{\gamma_{A,E}}{\gamma_{A,B}}2^{\frac{r_s}{\beta }}+1}$ and 
\begin{IEEEeqnarray}{rCl}\label{eqn_theta_D_1}
\Theta(D,1)=\frac{e^{-\tau}}{\gamma_{D^t,B}\:\eta}+\frac{e^{-\tau}\Bigg[\kappa\left(\eta+1\right)\Psi\left(\frac{\kappa+1}{\gamma_{D^t,E}}\right)+\left(\eta-\kappa\right)\Psi\left((\tau+\frac{1}{\gamma_{D^t,B}})(1+\frac{1}{\kappa})\right)\Bigg]}{\gamma_{D^t,B}\gamma_{D^t,E}\:\eta^2},
\end{IEEEeqnarray} 
where $\tau = \frac{2^{r_s}-1}{\gamma_{A,B}}$, $\kappa=\frac{2^{r_s}\gamma_{A,E}}{\gamma_{A,B}}$ and $\eta=\tau+\frac{1}{\gamma_{D^t,B}}-\frac{\kappa}{\gamma_{D^t,E}}$.

The ASC $\mathbf C_s$ of CUE $A$ is given by　
\begin{align}\label{eqn_case_cs}
\mathbf C_s = \frac{1}{\ln2 }\Big(p\,\Lambda(D)+\left(1-p\right)\beta\Lambda(\emptyset)\Big),
\end{align}
where $\Lambda(\emptyset)=\Psi\left(\frac{1}{\gamma_{A,B}}+\frac{1}{\gamma_{A,E}}\right)-\Psi\left(\frac{1}{\gamma_{A,B}}\right)$
and
\begin{IEEEeqnarray}{rCl}\label{eqn_lambda_D}
\Lambda(D)&=&\frac{1}{\frac{\gamma_{D^t,B}}{\gamma_{A,B}}-1}\Bigg[\frac{\Psi\left(\frac{\frac{\gamma_{A,E}}{\gamma_{A,B}}+1}{\gamma_{D^t,E}}\right)-\Psi\left(\frac{\frac{\gamma_{A,B}}{\gamma_{A,E}}+1}{\gamma_{D^t,B}}\right)}{\frac{\gamma_{A,B}\gamma_{D^t,E}}{\gamma_{A,E}\gamma_{D^t,B}}-1}+\frac{\Psi\left(\frac{1}{\gamma_{A,B}}+\frac{1}{\gamma_{A,E}}\right)-\Psi\left(\frac{\frac{\gamma_{A,E}}{\gamma_{A,B}}+1}{\gamma_{D^t,E}}\right)}{\frac{\gamma_{D^t,E}}{\gamma_{A,E}}-1}\nonumber\\
&&\ \ \ \ \ \ \ \ \ \ \ \ \ \ +\Psi\left(\frac{1}{\gamma_{A,B}}\right)-\Psi\left(\frac{1}{\gamma_{D^t,B}}\right)\Bigg].
\end{IEEEeqnarray}
\end{corollary}
\begin{IEEEproof}
The results follow from (\ref{theorem_sop}) and (\ref{theorem_asc}) in Theorem \ref{Theorem_CUE}, by letting $n=1$, $m=1$, $\vartheta=0$, $\varepsilon=p$, $f_{D}(x)=\frac{e^{-\frac{x}{\gamma_{D^t,B}}}}{\gamma_{D^t,B}}$ and $g_{ D}(x)=\frac{e^{-\frac{x}{\gamma_{D^t,E}}}}{\gamma_{D^t,E}}$, and then computing the involved integrals.
\end{IEEEproof}

Next, we give the following corollary regarding the analytical expressions for the OP and AC of D2D pair $D$.
\begin{corollary}\label{corollary_D2D}
Consider a cellular network with one base station $B$, one eavesdropper $E$, one CUE $A$ and one D2D pair $D$, the OP $\mathbf P_{o}$ of $D$ under the mode selection and spectrum partition schemes introduced in Section \ref{sec_mssp} is given by
\begin{align}\label{eqn_case_op} 
\mathbf P_{o}=1 - p\Omega(\emptyset,1)-(1-p)e^{-\frac{2^{\frac{r_t}{1-\beta}}-1}{\gamma_{D^{t},D^{r}}}},
\end{align}
where  $\Omega(\emptyset,1)=\frac{e^{-\frac{2^{r_t}\!-\!1}{\gamma_{D^{t},D^{ r}}}}}{\frac{\gamma_{A,D^{r}}}{\gamma_{D^{t},D^{ r}}}\left(2^{r_t}\!-\!1\right)+1}$.
The AC $\mathbf R$ of D2D pair $D$ is given by 
\begin{align}\label{eqn_case_r}
\mathbf R=\frac{1}{\ln 2}\left(\frac{p\Delta(\emptyset)}{\frac{\gamma_{A,D^{r}}}{\gamma_{D^{t},D^{r}}}-1}-(1-p)(1-\beta)\Psi\left(\frac{1}{\gamma_{D^{t},D^{r}}}\right)\right),
\end{align}
where  $\Delta(\emptyset)=\Psi\left(\frac{1}{\gamma_{D^{t},D^{r}}}\right)-\Psi\left(\frac{1}{\gamma_{A,D^{r}}}\right)$.
\end{corollary}
\begin{IEEEproof}
The results follow from (\ref{theorem_op}) and (\ref{theorem_ac}) in Theorem \ref{Theorem_D2D}, by letting $n=1$, $m=1$, $\vartheta=0$, $\varepsilon=p$ and $\mathbf D_d^{'}=\emptyset$.　
\end{IEEEproof} 

\subsection{Optimal Mode Selection and Spectrum Partition}
Based on the results in Corollaries \ref{corollary_CUE} and \ref{corollary_D2D}, we proceed to find the optimal settings of mode selection probability and spectrum partition factor by solving the optimization problems in Section \ref{sec_opt_formulation}. We first consider Problem P1, which is reduced to
\begin{align}\label{fn_OP1_case} 
(p_1^*,\beta_1^*) = \underset{p,\beta\in[0,1]}{\argmax}\,\, \mathcal U \left(\mathbf C_s,\mathbf R\right),
\end{align} 
where $\mathbf C_s$ and $\mathbf R$ are given in (\ref{eqn_case_cs}) and (\ref{eqn_case_r}), respectively. By applying Lemma \ref{lemma_OP1}, we can determine the  $p_1^*$ and $\beta_1^*$ in the following lemma.
\begin{lemma}\label{lemma_OP1_case}
For the considered case, the optimal mode selection probability $p_1^*$ and spectrum partition factor $\beta_1^*$ for Problem P1 are given as follows:　
\begin{enumerate} 
\item for $\mu+\nu>1$ and $w_c/\mu<w_d/\nu$,  where $\mu=\Lambda(D)/\Lambda(\emptyset)$ and $\nu=-\frac{\Delta(\emptyset)}{\left(\frac{\gamma_{A,D^{r}}}{\gamma_{D^{t},D^{r}}}-1\right)\Psi\left(\frac{1}{\gamma_{D^{t},D^{r}}}\right)}$, $(p_1^*,\beta_1^*)=\left(\min\left\{\frac{w_c}{1-\nu}, 1\right\},0\right)$; 
\item for $\mu+\nu>1$ and $w_c/\mu\geq w_d/\nu$, $(p_1^*, \beta_1^*)=\left(\min\left\{\frac{w_d}{1-\mu}, 1\right\}, 1\right)$, if $\mu<1$; otherwise $p_1^*=1$ and $\beta_1^*$ can be any value in $[0,1]$;　
\item for $\mu+\nu\leq1$, $(p_1^*, \beta_1^*)=(0, w_c)$.　 
\end{enumerate} 
\end{lemma}
\begin{IEEEproof}
See Appendix \ref{proof_lemma_OP1_case}.
\end{IEEEproof}
Notice that the $\mu$ (resp. $\nu$) in Lemma \ref{lemma_OP1_case} is the ratio of the ASC for CUE $A$ (resp. the AC for D2D pair $D$) when  $D$ operates in the underlay mode to that when $D$ operates in the overlay mode with full spectrum usage, i.e, $\beta=1$ (resp. $\beta=0$). We can interpret $\mu$ and $\nu$ as the underlay rate gains of the CUE and the D2D pair, and their reciprocals $1/\mu$ and $1/\nu$ as the overlay rate gains. From cases 1) and 2), we can see that when the system underlay gain (i.e., $\mu+\nu$) is greater than $1$, to achieve the optimal system rate performance, on the one hand, the D2D pair is encouraged to reuse the spectrum resource of the CUE with a certain probability. On the other hand, when the D2D pair chooses the overlay mode, the base station allocates all its spectrum resource to the D2D pair (i.e., $\beta_1^*=0$) if the weighted overlay gain of the CUE is less than that of the D2D pair; otherwise, the base station allocates all its spectrum to the CUE (i.e., $\beta_1^*=1$). From case 3), we can see that when the system underlay gain is less than $1$, to optimize the optimal system rate performance, the base station disables the spectrum reuse between the CUE and the D2D pair, and sets the spectrum partition factor as the weight assigned to the utility of the CUE. Special attention needs to be paid to the optimal solutions with $p_1^*=1$. In this case, the D2D pair reuses the spectrum of the CUE and all the spectrum is allocated to the CUE, so no spectrum partition is needed and the optimal spectrum partition factor $\beta_1^*$ can be any value in $[0,1]$.
 
Next, we consider Problem P2, which is reduced to 
\begin{align}\label{fn_OP2_case}
(p_2^*, \beta_2^*) = \underset{p,\beta\in[0,1]}{\argmax}\,\, -\mathcal U \left(\mathbf P_{so},\mathbf P_{o}\right),
\end{align} 
where $\mathbf P_{so}$ and $\mathbf P_{o}$ are given by (\ref{eqn_case_sop}) and (\ref{eqn_case_op}), respectively. According to the discussions in Section \ref{sec_opt_formulation}, a two-dimensional search over $(p, \beta)$ is usually required to find the  $p_2^*$ and  $\beta_2^*$. However, for this simple case, we find that  $p_2^*$ is either $p_2^*=0$ or $p_2^*=1$ due to the convexity of the objective function in terms of $p$. Based on this property, we can find the $p_2^*$ and $\beta_2^*$ for this problem. Before giving the main results, we first define two functions in terms of $\beta$, i.e., $\hat{\mu}(\beta)=(1-\Theta(\emptyset, \beta))/(1-\Theta(D,1))$ and $\hat{\nu}(\beta)=(1-e^{-\frac{2^{\frac{r_t}{1-\beta}}-1}{\gamma_{D^{t},D^{r}}}})/(1-\Omega(\emptyset,1))$. In addition, we define a set $\varrho=\{\beta|\hat{\mu}(\beta)^{w_c}\hat{\nu}(\beta)^{w_d}<1\}$. With the help of these definitions, we can give the following lemma about $p_2^*$ and $\beta_2^*$.
 
\begin{lemma}\label{lemma_OP2_case}
For the considered case, the optimal mode selection probability $p_2^*$ and spectrum partition factor $\beta_2^*$ for Problem P2 are given by
 $(p_2^*, \beta_2^*)=(0, \hat{\beta}_2)$, where
\begin{align}
\hat{\beta}_2=\underset{\beta \in \varrho}{\argmax}\,\, -\mathcal U \left(1-\Theta(\emptyset,\beta),1-e^{-\frac{2^{\frac{r_t}{1-\beta}}\!-\!1}{\gamma_{D^{t},D^{r}}}},\right)
\end{align}
if $\varrho\neq \emptyset$, otherwise $p_2^*=1$ and $\beta_2^*$ can be any value in $[0,1]$.
\end{lemma}
\begin{IEEEproof}
We first take the second partial derivative of the objective function with respect to $p$, which is,
\begin{align}
-\frac{\partial^2 \mathcal U}{\partial p^2}=\frac{w_c}{\left(p+\frac{1}{1/\hat{\mu}(\beta)-1}\right)^2}+\frac{w_d}{\left(p+\frac{1}{1/\hat{\nu}(\beta)-1}\right)^2}>0.
\end{align}
We can see that the objective function is a convex function of $p$. Thus, the maximum can only be achieved at either $p=0$ or $p=1$. For $\beta \in \varrho$, we have $-\mathcal U_{p=0}>-\mathcal U_{p=1}$ and the optimal $p$ is $p_2^*=0$. The optimal $\beta$ in this case is given by $\beta_2^*=\hat{\beta}_2$. For $\beta \notin \varrho$, the optimal $p$ is $p_2^*=1$. The optimal $\beta$ can be any $\beta$  in the complement of $\varrho$, as the objective function is independent of $\beta$ for $p=1$. If $\varrho \neq \emptyset$, we have $-\mathcal U_{p=0,\beta=\hat{\beta}_2}>-\mathcal U_{p=1,\beta=\hat{\beta}_2}=-\mathcal U_{p=1,\beta\notin \varrho}$, and thus $\beta_2^*=\hat{\beta}_2$ and $p_2^*=0$. Otherwise,  $p_2^*=1$ and $\beta_2^*$ can be any value in $[0,1]$. 
\end{IEEEproof}
Notice that the $\hat{\mu}(\beta)$ (resp. $\hat{\nu}(\beta)$) in Lemma \ref{lemma_OP2_case} is the ratio of the SOP for CUE $A$ (resp. the OP for D2D pair $D$) when $D$ operates in the overlay mode to that when $D$ operates in the underlay mode with a fixed $\beta$. Thus, the $\hat{\mu}(\beta)$ (resp. $\hat{\nu}(\beta)$) can be interpreted as the underlay security gain of the CUE (resp. the underlay reliability gain of the D2D pair), and its reciprocal $1/\hat{\mu}(\beta)$ (resp. $1/\hat{\nu}(\beta)$) as the corresponding overlay gains. We can see from Lemma \ref{lemma_OP2_case} that if there exists at least one spectrum partition factor $\beta$ such that $\hat{\mu}^{w_c}(\beta)\hat{\nu}^{w_d}(\beta)<1$, the optimal mode selection probability is $p_2^*=0$. In this case, the D2D pair should choose the overlay mode to minimize the system outage probability. 

\section{Numerical Results for Case Study} \label{sec_num_dis}  
In this section, we first provide simulation results to validate the analytical expressions of SOP, ASC, OP and AC for the case study. We then explore how these performances vary with the mode selection probability $p$, spectrum partition factor $\beta$ and other system settings. Finally, we demonstrate the feasibility of our approach to find the optimal settings of mode selection probability and spectrum partition factor. 
 \begin{table}[!t]
\renewcommand{\arraystretch}{0.9}
\caption{Simulation Parameters}
\label{tb_pmts}
\centering
\begin{tabular}{|l|l|}
\hline
\bfseries Parameter &  \bfseries Value \\
\hline
Cell radius  & $500$ m\\
\hline
Location of base station & ($0$ m, $0$ m)\\
\hline
Total Bandwidth $W$ & 1 MHz\\
\hline
Noise spectral density & -174 dBm/Hz \\
\hline
Path loss exponent $\alpha$ & 4 \\
\hline
Small-scale fading & Rayleigh fading \\
\hline
Transmit power of cellular user $P_A$ & 23 dBm \\
\hline
Transmit power of D2D user $P_D$ & 20 dBm  \\
\hline
\end{tabular}
\end{table}
\subsection{Simulation Settings and Model Validation}
We developed a dedicated simulator in C++ for the case study to simulate the message transmission processes of both the CUE and D2D pair, which is now available at \cite{Mdval5}. We consider an isolated cellular cell with a radius of $500$ m. The base station $B$ is located at the center ($0$ m, $0$ m). The simulation parameters are summarized in Table \ref{tb_pmts}. To verify the accuracy of our theoretical analysis, we compare the simulated and theoretical values of the SOP, ASC, OP and AC. Each simulated value is calculated as the average value of $10^2$ batches of simulation results. In each batch, $10^6$ random and independent simulations are conducted and the corresponding SOP (resp. OP) is calculated as the ratio of the number of simulations with secrecy outage (resp. outage) to the total number of simulations $10^6$. Similarly, the ASC and AC in each patch are calculated as the average value of the ASC and AC of $10^6$ simulations, respectively.

\begin{table}[!t]
\renewcommand{\arraystretch}{0.9}
\caption{SOP and ASC validation for CUE at ($100$ m, $100$ m), $r_s = 0.1$ Mbits/s, $p=0.5$, $\beta = 0.5$, Simulated/Theoretical}
\label{tb_mv_sop_asc}
\centering
\begin{tabular}{ |c|c|c|c| }
\hline
$E$ & $D^t$  & SOP & ASC \\ \hline
\multirow{3}{*}{($0$, $100$)} & ($100$, $0$) & $0.851151\pm 0.004/0.851236$ & $0.161597\pm0.004/0.1614$\\
\cline{2-4}
 & ($0$, $200$) & $0.559092\pm 0.004/0.559389$ & $1.03312\pm0.006/1.03242$\\
\cline{2-4}
 & ($50$, $100$) & $0.554693\pm0.003/0.55475$ &$0.582294\pm0.008/0.58243$\\ 
\hline
\multirow{3}{*}{($0$, $200$)} & ($0$, $300$) & $0.290782\pm0.007/0.290804$ & $2.51383\pm0.02/2.51353$\\
\cline{2-4}
 & ($50$, $200$) & $0.279723\pm0.004/0.279718$ &$1.96078\pm0.02/1.95986$\\
\cline{2-4}
 & ($100$, $200$) & $0.321149\pm0.003/0.320976$&$1.78978\pm0.01/1.79134$\\ 
\hline
\multirow{3}{*}{($0$, $300$)} & ($0$, $100$) & $0.421326\pm0.006/0.421194$ & $0.915345\pm0.02/0.91561$\\
\cline{2-4}
 & ($0$, $200$) & $0.101184\pm0.002/0.101194$ & $2.28495\pm0.02/2.28545$\\
\cline{2-4}
 & ($100$, $300$) & $0.0830492\pm0.002/0.0829538$ &$3.43883\pm0.02/3.43743$\\ 
\hline
\end{tabular}
\end{table}

\begin{table}[!t]
\renewcommand{\arraystretch}{0.9}
\caption{OP and AC validation for CUE at ($100$ m, $100$ m), $r_t=0.5$ Mbits/s, $p=0.5$, $\beta = 0.5$, Simulated/Theoretical}
\label{tb_mv_op_ac}
\centering
\begin{tabular}{ |c|c|c|c| }
\hline 
$D^t$ & $D^r$  & OP & AC \\ \hline
\multirow{3}{*}{($100$, $0$)} & ($100$, $50$) & $0.226646\pm 0.004/0.226541$ & $5.77726\pm0.05/5.77642$\\
\cline{2-4}
 & ($150$, $50$) & $0.226472\pm 0.004/0.226542$ & $5.27576\pm0.04/5.27642$\\
\cline{2-4}
 & ($150$, $0$) & $0.0159588\pm0.001/0.0160373$ &$7.25613\pm0.03/7.25675$\\ 
\hline
\multirow{3}{*}{($200$, $0$)} &($200$, $50$) & $0.0160605\pm0.001/0.0160373$ & $7.25671\pm0.02/7.25675$\\
\cline{2-4}
 & ($150$, $100$) & $0.476906\pm0.005/0.476973$&$4.1746\pm0.04/4.17307$\\
\cline{2-4}
 & ($250$, $0$) & $0.0024383\pm0.0004/0.00243919$&$8.51556\pm0.03/8.51502$\\ 
\hline
\multirow{3}{*}{($300$, $0$)} & ($300$, $50$) & $0.0014218\pm0.0003/0.00142934$ & $8.8886\pm0.03/8.8886$\\
\cline{2-4}
 & ($300$, $100$) & $0.024502\pm0.001/0.0246167$ & $5.99154\pm0.02/5.99048$\\
\cline{2-4}
 & ($350$, $0$) & $0.0005022\pm0.0002/0.000492216$ &$9.94441\pm0.02/9.64406$\\ 
\hline
\end{tabular}
\end{table}
For the validation of SOP and ASC, we place the CUE $A$ at ($100$ m, $100$ m) and set the target secrecy rate as $r_s = 0.1$ Mbits/s, the mode selection probability as $p=0.5$ and the spectrum partition factor as $\beta = 0.5$. We consider three different cases of the location of the eavesdropper $E$, i.e., ($0$ m, $100$ m), ($0$ m, $200$ m) and ($0$ m, $300$ m). For each case, three different locations of the D2D transmitter $D^t$ have been considered. The simulated and theoretical values are summarized in Table \ref{tb_mv_sop_asc}. For the validation of OP and AC, we place $A$ at ($100$ m, $100$ m) and set $p = 0.5$, $\beta = 0.5$ and the target rate as $r_t = 0.5$ Mbits/s. We consider three different cases of the location of $D^t$, i.e., ($100$ m, $0$ m), ($200$ m, $0$ m) and ($300$ m, $0$ m). For each case, three different locations of the D2D receiver $D^r$ have been examined. The simulated and theoretical values are summarized in Table \ref{tb_mv_op_ac}. We can see from Table \ref{tb_mv_sop_asc} and Table \ref{tb_mv_op_ac} that the simulated values match proficiently with the theoretical ones, which indicates that our theoretical framework is efficient to model the SOP and ASC of CUEs as well as the OP and AC of D2D pairs under the considered mode selection and spectrum partition schemes. 

\subsection {Performance Evaluation}
\begin{figure}[!t]
\centering
\subfloat[SOP vs. $y$ coordinate of $D^t$.]{\includegraphics[width=2.8in]{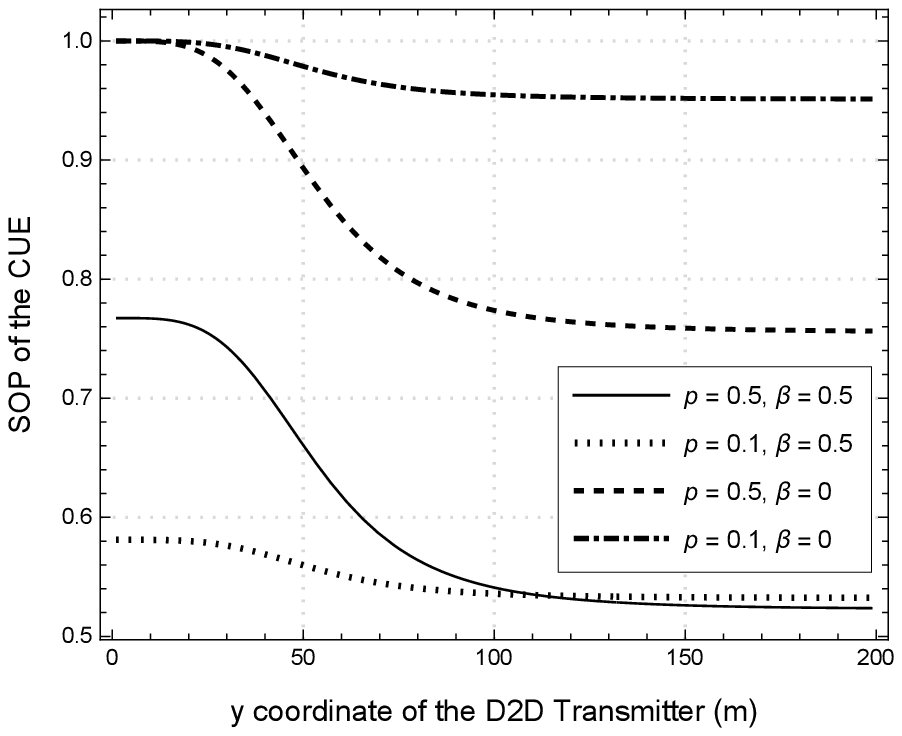}
\label{fig_sop_p_rho}}
\hfil
\subfloat[ASC vs. $y$ coordinate of $D^t$.]{\includegraphics[width=2.8in]{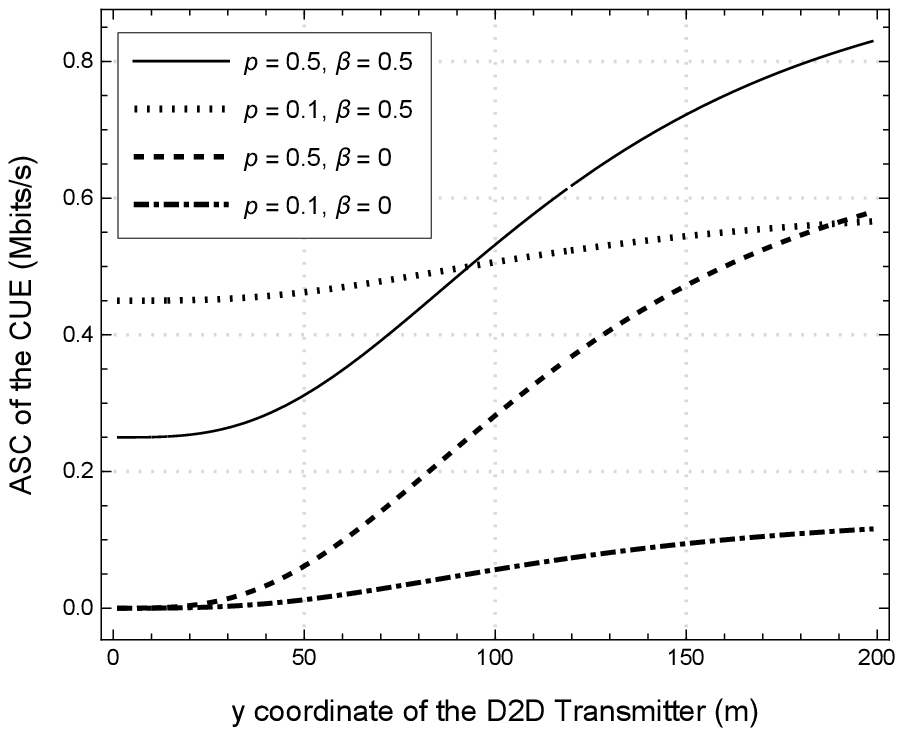}
\label{fig_asc_p_rho}}
\caption{Impacts of $p$, $\beta$ and location of $D^t$ on the CUE security performances. Spatial distribution of network users: $A$ = ($100$ m, $100$m), $E$ = ($0$ m, $200$ m) and $D^t$ = ($0$ m, $[1-199]$ m).}
\label{fig_cue_p_rho}
\end{figure}

We now investigate how the mode selection probability $p$, spectrum partition factor $\beta$ and the location of the D2D transmitter $D^t$ affect the SOP and ASC performances of the CUE $A$. We consider a scenario where $A$ is located at ($100$ m, $100$ m) and $E$ is located at ($0$ m, $200$ m). We fix the $x$ coordinate of $D^t$ as $0$ m and vary its $y$ coordinate from $1$ m to $199$ m. For this scenario, we show in Fig. \ref{fig_cue_p_rho} how the SOP and ASC of the CUE vary with the $y$ coordinate of $D^t$ under different settings of $p$ (i.e., $0.1$ and $0.5$) and $\beta$ (i.e., $0$ and $0.5$) for $r_s=0.1$ Mbits/s.  We can see from Fig. \ref{fig_sop_p_rho} that the SOP decreases as the $y$ coordinate of $D^t$ increases. This implies that a larger ratio of the distances $d_{D^t,B}$ to $d_{D^t,E}$ (i.e., $d_{D^t,B}/d_{D^t,E}$) can achieve a lower SOP. We can also see from Fig. \ref{fig_sop_p_rho} that the SOP always decreases as $\beta$ increases and this trend is independent of both $p$ and the location of $D^t$. On the contrary, however, the behavior of SOP versus $p$ depends on both $\beta$ and the location of $D^t$. For example, for the case of $\beta=0$, the SOP decreases as $p$ increases for all $y$ coordinates of $D^t$, while for the case of $\beta=0.5$, the SOP increases as $p$ increases when the $y$ coordinate of $D^t$ is less than a threshold (about $110$ m in Fig.\ref{fig_sop_p_rho}) but decreases as $p$ increases when the $y$ coordinate of $D^t$ is larger than the threshold. This is because that, as can be seen from (\ref{eqn_case_sop}), the condition for SOP decreasing as $p$ increases is $\hat{\mu}(\beta)>1$, i.e., the SOP achieved when the D2D pair chooses the underlay mode is less than that when the D2D pair chooses the overlay mode. 

Regarding the impacts of $p$, $\beta$ and the location $D^t$ on the ASC of the CUE, we can see from Fig. \ref{fig_asc_p_rho} that the ASC increases as the $y$ coordinate of $D^t$ increases, which implies that a larger distance ratio $d_{D^t,B}/d_{D^t,E}$ is also beneficial for increasing the ASC. Similar to the behaviors of the SOP versus $p$ and $\beta$, it can also be observed from Fig. \ref{fig_asc_p_rho} that the ASC always increases as $\beta$ increases, while the ASC increases with $p$ only if the ASC achieved when the D2D pair chooses the underlay mode is larger than that when the D2D pair chooses the overlay mode (i.e., $\mu>\beta$).

\begin{figure}[!t]
\centering
\subfloat[OP vs. $\gamma_{A,D^{r}}/\gamma_{D^{t},D^{r}}$.]{\includegraphics[width=2.8in]{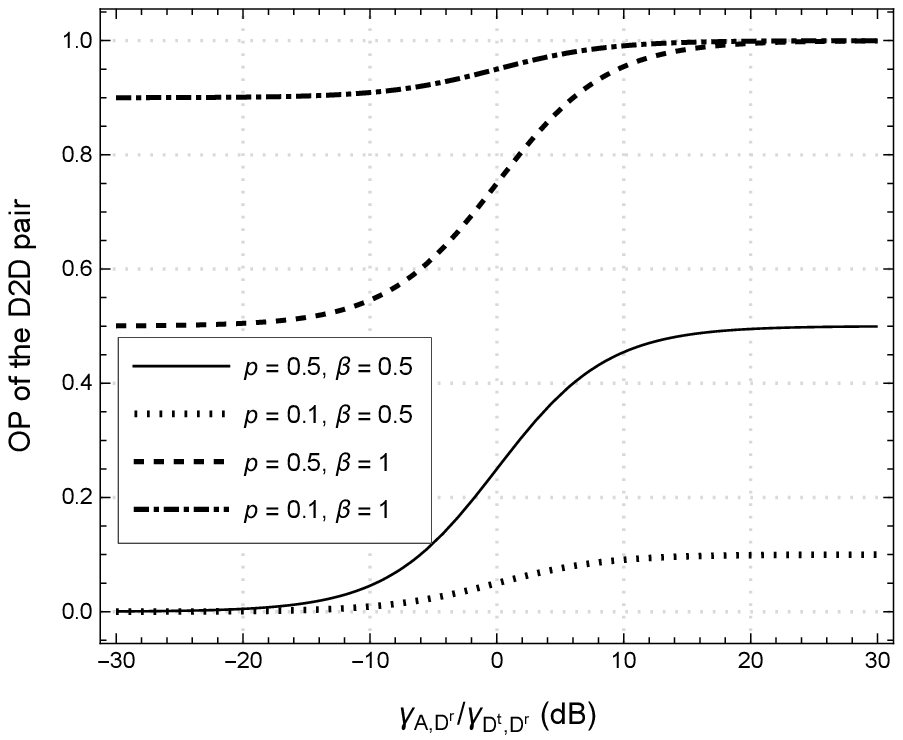}
\label{fig_op_p_rho}}
\hfil
\subfloat[AC vs. $\gamma_{A,D^{r}}/\gamma_{D^{t},D^{r}}$.]{\includegraphics[width=2.8in]{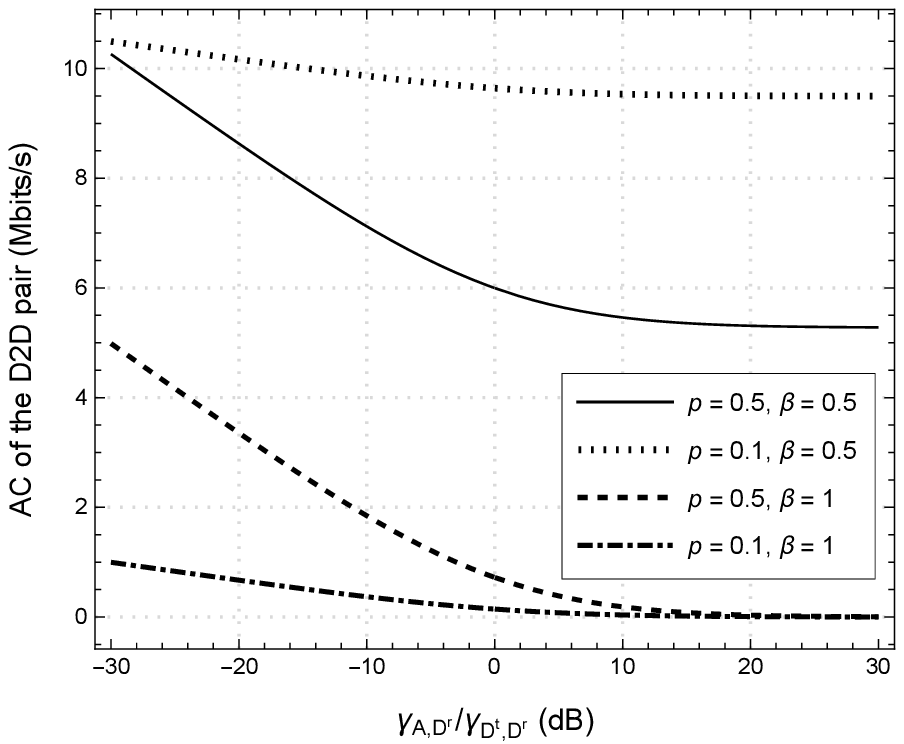}
\label{fig_ac_p_rho}}
\caption{Impact of  $p$, $\beta$ and $\gamma_{A,D^{r}}/\gamma_{D^{t},D^{r}}$ on D2D performances with $r_t=1$ Mbits/s and $d_{D^t,D^r}=100$ m.}
\label{fig_d2d_p_rho}
\end{figure}

We now examine the impacts of $p$, $\beta$  and the ratio of the average SNR $\gamma_{A,D^{r}}$ to the average SNR $\gamma_{D^{t},D^{r}}$ (i.e., $\gamma_{A,D^{r}}/\gamma_{D^{t},D^{r}}$) on the performances of the D2D pair. For the scenario with $r_t=1$ Mbits/s and $d_{D^t,D^r}=100$ m, Fig. \ref{fig_d2d_p_rho} illustrates the OP and AC of the D2D pair versus $\gamma_{A,D^{r}}/\gamma_{D^{t},D^{r}}$ under different settings of $p$ and $\beta$. We can see from Fig. \ref{fig_op_p_rho} that for a given SNR $\gamma_{D^{t},D^{r}}$, the OP of the D2D pair increases as the SNR $\gamma_{A,D^{r}}$ increases. This is because that, when the D2D pair selects the underlay mode, more interference is generated by the CUE at the D2D receiver $D^r$, resulting in a lower communication rate and thus a larger outage probability. We can also see from Fig. \ref{fig_op_p_rho} that the OP always increases as $\beta$ increases, which is regardless of the values of $p$ and the SNR ratio $\gamma_{A,D^{r}}/\gamma_{D^{t},D^{r}}$. In contrast, we can see that the behavior of OP versus $p$ depends on both $\beta$ and $\gamma_{A,D^{r}}/\gamma_{D^{t},D^{r}}$. For example, the OP increases as $p$ increases for $\beta=0.5$, while it decreases as $p$ increases for $\beta=1$. This is due to the reason that the OP decreases as $p$ increases if the OP achieved when the D2D pair operates in the underlay mode is less than that when the D2D pair operates in the overlay mode, i.e., $\hat{\nu}(\beta)>1$ as can be inferred from (\ref{eqn_case_op}). 

Similar behaviors of the AC versus $p$, $\beta$ and $\gamma_{A,D^{r}}/\gamma_{D^{t},D^{r}}$ can also be observed from Fig. \ref{fig_ac_p_rho}. These observations are: 1) the AC decreases as the ratio $\gamma_{A,D^{r}}/\gamma_{D^{t},D^{r}}$ increases for a given $\gamma_{D^{t},D^{r}}$ due to the more interference at $D^r$ generated by operating in the underlay mode; 2) the AC always decreases as $\beta$ increases, while it decreases as $p$ increases only if the AC achieved when the D2D pair operates in the underlay mode is less than that when the D2D pair operates in the overlay mode (i.e., $\nu < 1-\beta$). 
\begin{figure*}[!t]
\centering
\subfloat[Case 1): $\mu=1.09974$, $\nu=0.187658$, $w_c=0.4$ and $w_d=0.6$.]{\includegraphics[width=2.8in]{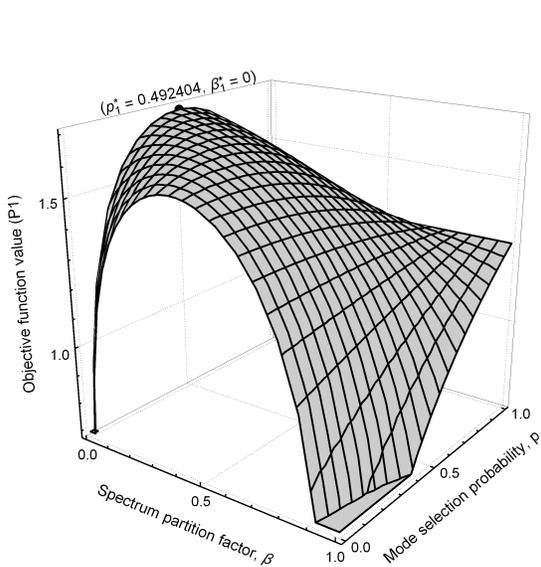}
\label{fig_op1_case1}}
\hfil
\subfloat[Case 2): $\mu=0.770184$, $\nu=0.28467$, $w_c=0.9$ and $w_d=0.1$.]{\includegraphics[width=2.8in]{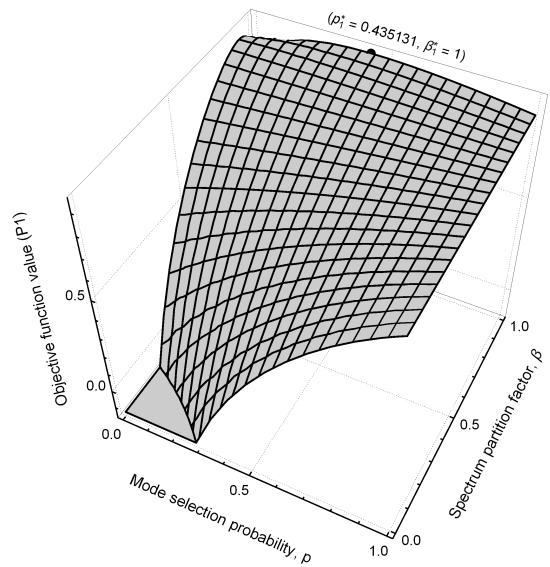}
\label{fig_op1_case2_1}}
\vfil
\subfloat[Case 2): $\mu=1.09974$, $\nu=0.187658$, $w_c=0.9$ and $w_d=0.1$.]{\includegraphics[width=2.8in]{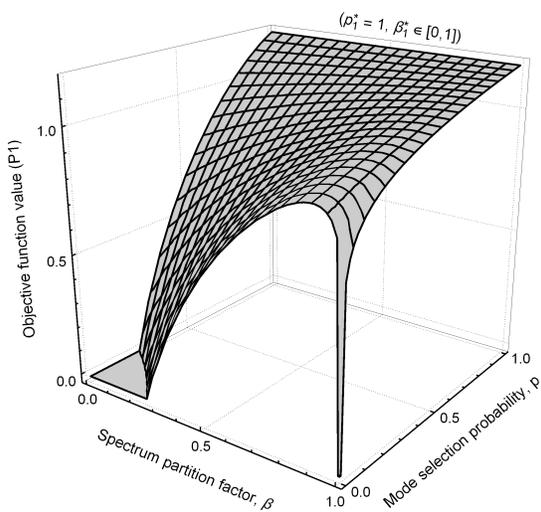}
\label{fig_op1_case_2}}
\hfil
\subfloat[Case 3): $\mu=0.552902$, $\nu=0.126527$, $w_c=0.4$ and $w_d=0.6$.]{\includegraphics[width=2.8in]{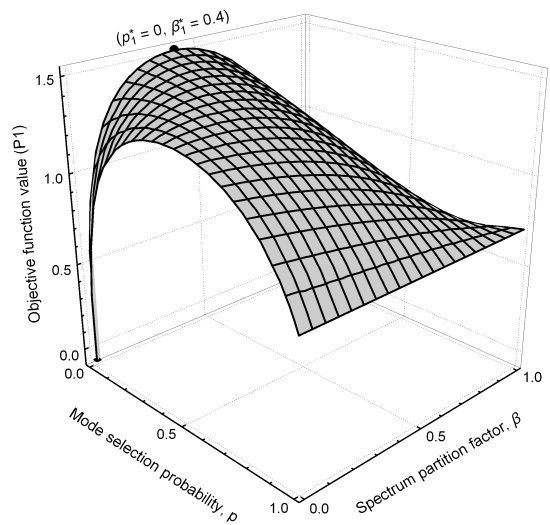}
\label{fig_op1_case_3}}
\caption{Optimal mode selection probability $p_1^*$ and spectrum partition factor $\beta_1^*$ of Problem P1. Spatial distribution of network users: $A$ = ($100$ m, $100$ m) and  $E$ = ($0$ m, $300$ m); $D^t$ = ($0$ m, $200$ m) and $D^r$ = ($50$ m, $200$ m) for (a) and (c), $D^t$ = ($0$ m, $170$ m) and $D^r$ = ($-50$ m, $170$ m) for (b), $D^t$ = ($0$ m, $150$ m) and $D^r$ = ($50$ m, $200$ m) for (d). }
\label{fig_op1}
\end{figure*}

\subsection{Optimal $p$ and $\beta$}
We first study the optimal mode selection probability $p_1^*$ and spectrum partition factor $\beta_1^*$ of Problem P1. We show in Fig. \ref{fig_op1} the objective function of Problem P1 versus $p$ and $\beta$ for $A$ = ($100$ m, $100$ m) and $E$ = ($0$ m, $300$ m).  Fig. \ref{fig_op1} includes four sub-figures with different settings of weights $w_c$ and $w_d$ and locations of $D^t$ and $D^r$, which correspond to different cases in Lemma \ref{lemma_OP1_case}. Fig. \ref{fig_op1_case1} corresponds to case 1) with $\mu=1.09974$, $\nu=0.187658$, $w_c=0.4$ and $w_d=0.6$ under the scenario of $D^t$ = ($0$ m, $200$ m) and $D^r$ = ($50$ m, $200$ m). We can see from Fig. \ref{fig_op1_case1} that the maximum of the objective function is achieved at $(p_1^*=0.492404, \beta_1^*=0)$, which matches the solution of $(p_1^*=w_c/(1-\nu), \beta_1^*=0)$ in case 1) of Lemma \ref{lemma_OP1_case}. Fig. \ref{fig_op1_case2_1} corresponds to case 2) with $\mu=0.770184$, $\nu=0.28467$, $w_c=0.9$ and $w_d=0.1$ under the scenario of $D^t$ = ($0$ m, $170$ m) and $D^r$ = ($-50$ m, $170$ m). It can be seen from Fig. \ref{fig_op1_case2_1} that the maximum of the objective function is achieved at $(p_1^*=0.435131, \beta_1^*=1)$, which matches the optimal solution of $(p_1^*=w_d/(1-\mu), \beta_1^*=1)$ for case 2) with $\mu<1$ in Lemma \ref{lemma_OP1_case}. To illustrate the solutions of case 2) with $\mu>1$, we consider the same locations of $D^t$ and $D^r$ (i.e., $\mu=1.09974$, $\nu=0.187658$) as for case 1) and show the results in Fig. \ref{fig_op1_case_2} for $w_c=0.9$ and $w_d=0.1$. As can be seen from Fig. \ref{fig_op1_case_2}, the optimal $p_1^*$ is $p_1^*=1$ and optimal $\beta_1^*$ can be any value in $[0,1]$, which verifies the solution for case 2) with $\mu>1$. Finally, Fig. \ref{fig_op1_case_3} illustrates the solution in case 3) with $\mu=0.552902$, $\nu=0.126527$, $w_c=0.4$ and $w_d=0.6$ for the scenario of $D^t$ = ($0$ m, $150$ m) and $D^r$ = ($50$ m, $200$ m). Fig. \ref{fig_op1_case_3} shows that the optimal solution is $(p_1^*=0, \beta_1^*=0.4)$, which agrees with the solution of $(p_1^*=0, \beta_1^*=w_c)$ for case 3) in Lemma \ref{lemma_OP1_case}. Therefore, the above figures demonstrate the feasibility of our approach for determining the  $p_1^*$ and $\beta_1^*$ of Problem P1. 

\begin{figure}[!t]
\centering
\subfloat[$w_c=0.4$ and $w_d=0.6$.]{\includegraphics[width=2.8in]{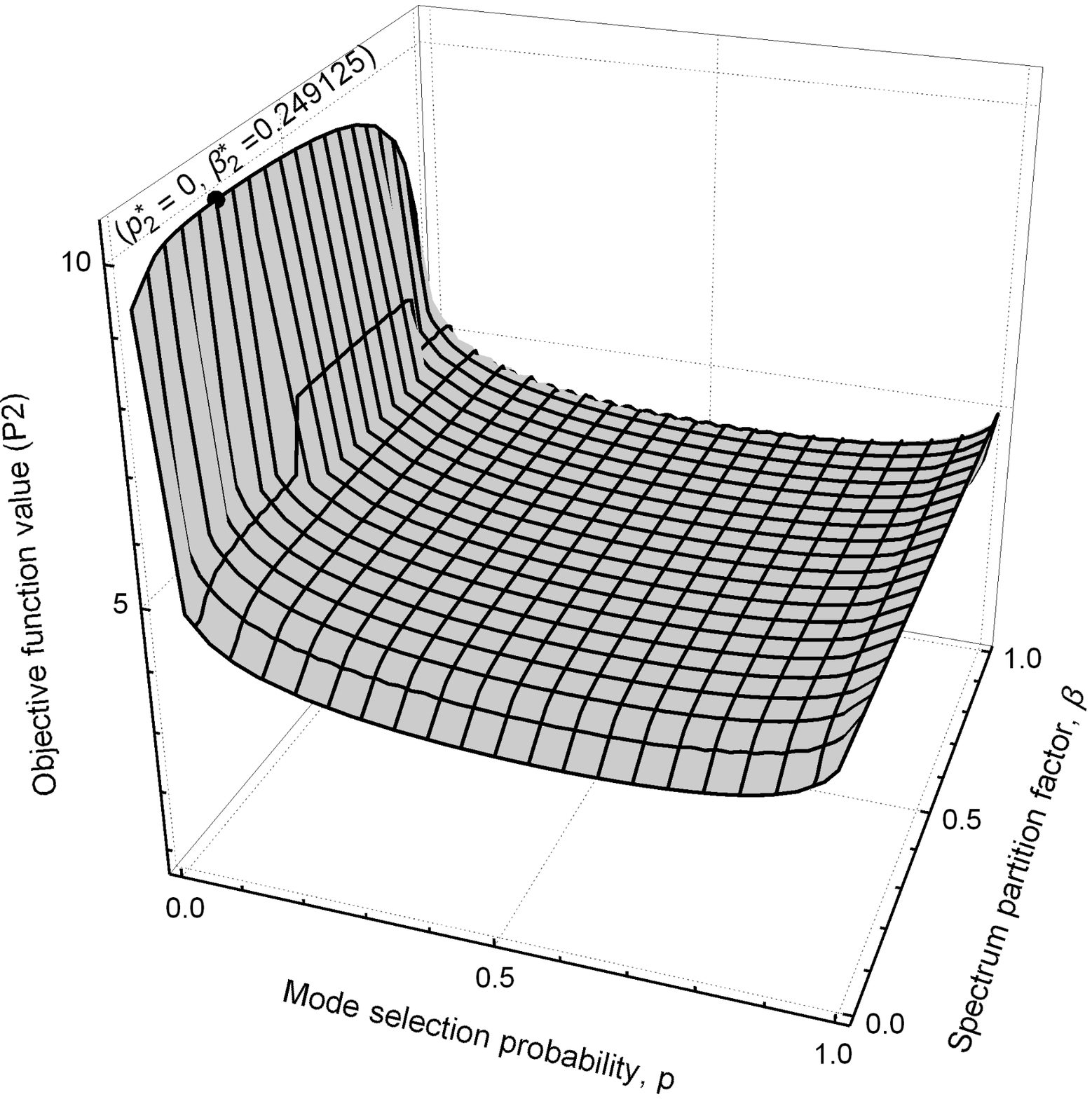}
\label{fig_op2_case1}}
\hfil
\subfloat[$w_c=0.9$ and $w_d=0.1$.]{\includegraphics[width=2.8in]{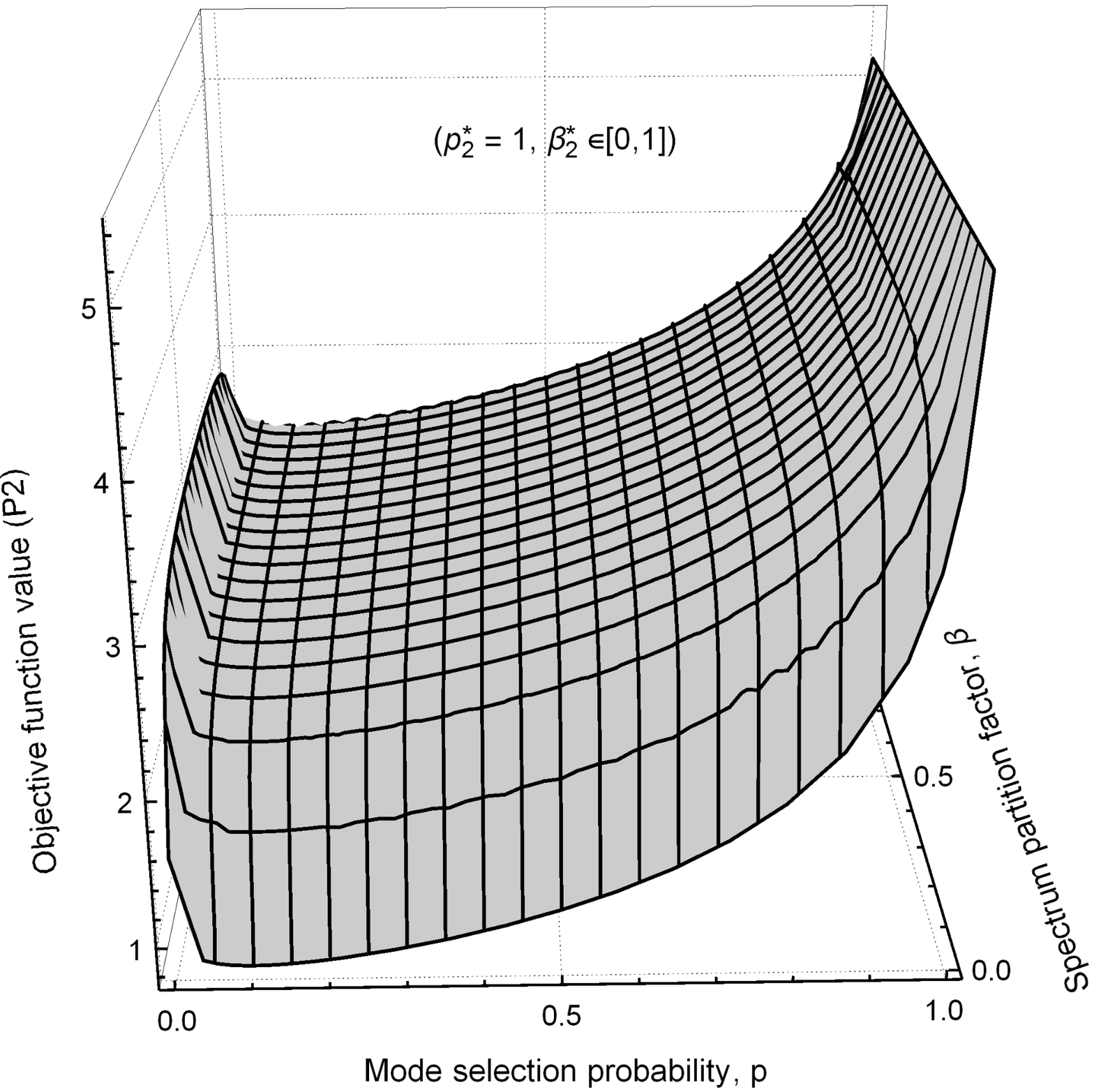}
\label{fig_op2_case2}}
\caption{Optimal mode selection probability $p_2^*$ and spectrum partition factor $\beta_2^*$ of Problem P2. Spatial distribution of network users: $A$ = ($100$ m, $100$ m), $E$ = ($0$ m, $300$ m), $D^t$ = ($0$ m, $250$ m) and $D^r$ = ($50$ m, $250$ m). }
\label{fig_op2}
\end{figure}

Next, we explore the optimal mode selection probability $p_2^*$ and spectrum partition factor $\beta_2^*$ of Problem P2. For the scenario of $A$ = ($100$ m, $100$ m), $E$ = ($0$ m, $300$ m), $D^t$ = ($0$ m, $250$ m) and $D^r$ = ($50$ m, $250$ m), Fig. \ref{fig_op2_case1} and Fig. \ref{fig_op2_case2} illustrate the objective function of Problem P2 versus $p$ and $\beta$ for the settings of $w_c=0.4$, $w_d=0.6$ and $w_c=0.9$, $w_d=0.1$, respectively. We can see from Fig. \ref{fig_op2_case1} and Fig. \ref{fig_op2_case2} that for a given $\beta$, the objective function of Problem P2 is a convex function of $p$, and thus the optimal $p_2^*$ that maximizes the objective function can only be either $p_2^*=0$ or $p_2^*=1$. For the settings of $w_c=0.4$ and $w_d=0.6$, we can see from Fig. \ref{fig_op2_case1} that there exist at least one $\beta$ such that the value of the objective function at $p=0$ is greater than that at $p=1$ (i.e., the set $\varrho$ is not empty). Thus, the optimal $p_2^*$ in this case is $p_2^*=0$. We can also verify that the optimal $\beta_2^*$ shown in Fig. \ref{fig_op2_case1} agrees with the optimal $\beta_2^*$ that maximizes the objective function for $p_2^*=0$. However, for the settings of $w_c=0.9$ and $w_d=0.1$, we can see from Fig. \ref{fig_op2_case2} that the value of the objective function at $p=0$ is always smaller than that at $p=1$ (i.e., the set $\varrho$ is empty). Thus, the optimal $p_2^*$ in this case is $p_2^*=1$ and the optimal $\beta_2^*$ can be any value in $[0,1]$. These observations from \ref{fig_op2_case1} and Fig. \ref{fig_op2_case2} match the results in Lemma \ref{lemma_OP2_case} and demonstrate the feasibility of our approach for determining the $p_2^*$ and $\beta_2^*$ of Problem P2.

\section{Conclusions} \label{sec_con} 
This paper investigated the mode selection and spectrum partition issues in cellular networks with inband D2D communication from the physical layer security (PLS) perspective. In particular, we developed a theoretical framework for the performance modeling and optimization of both cellular users (CUEs) and D2D pairs. The results based on a case study indicated that the PLS performances of the CUE and D2D pair can be flexibly controlled by mode selection and spectrum partition. For example, we can reduce (resp. increase) the secrecy outage probability (resp. average secrecy capacity) of the CUE by allocating more spectrum to the CUE, but such PLS performance improvement of CUE usually comes with the cost of an increased outage probability (resp. reduced average capacity) of the D2D pair. On the other hand, the mode selection of D2D pair has a complicated impact on the PLS performance of CUE and reliable communication performance of D2D pair, which is heavily dependent on the setting of spectrum partition and the spatial distribution of network users.

\appendices
\section{Proof of Theorem \ref{Theorem_CUE}}\label{app_theorem_cue} 
We first prove the SOP result in (\ref{theorem_sop}). Following from the definition of SOP in (\ref{sop_def}), we have 
\begin{align}
\mathbf P_{so}^{\,i}=\mathbb P\left[\frac{1+\frac{\mathrm{SNR}_{A_i,B}}{I_B+1}}{1+\frac{\mathrm{SNR}_{A_i,E}}{I_E+1}}<2^{\frac{nr_s}{\beta }}\right].
\end{align}
As $\mathrm{SNR}_{A_i,B}$ and $\mathrm{SNR}_{A_i,E}$ are independent exponentially distributed random variables, we have
\begin{align}\label{eqn_exp}
\mathbf P_{so}^{\,i}=1-\mathbb E\left[\frac{e^{-\frac{(2^{\frac{nr_s}{\beta}}-1) (I_B+1)}{\gamma_{A_i,B}}}}{\frac{\gamma_{A_i,E}}{\gamma_{A_i,B}}\frac{I_B+1}{I_E+1}2^{\frac{nr_s}{\beta}}+1}\right],
\end{align}
where the expectation is with respect to $I_B$ and $I_E$. As both $I_B$ and $I_E$ depend on $\mathbf D_u$, we applying the law of total expectation in terms of $\mathbf D_u$ to change (\ref{eqn_exp}) to 
\begin{align}\label{term_T}
\mathbf P_{so}^{\,i}\!=\!1\!-\!\!\!\!\!\sum_{\mathbf D_u\in 2^{\mathcal D}}\!\!\!\underbrace{\mathbb E\!\!\left[\frac{e^{-\frac{(2^{\frac{nr_s}{\beta }}-1) (I_B+1)}{\gamma_{A_i,B}}}}{\frac{\gamma_{A_i,E}}{\gamma_{A_i,B}}\frac{I_B+1}{I_E+1}2^{\frac{nr_s}{\beta }}\!+\!1}\Bigg|\mathbf D_u\right]\!\mathbb P\!\!\left(\mathbf D_u\right)}_{T},
\end{align}
where $\mathbb P\left(\mathbf D_u\right)$ denotes the probability $\mathbb P\left(\mathcal D_{d}^{i}=\mathbf D_u\right)$. For the special case where all the other $m-|\mathbf D_u|$ pairs also select the underlay mode (i.e., $\mathcal D_u=\mathcal D$ and $\mathcal D_o=\emptyset$), we have $\beta=1$ and $\mathbb P\left(\mathbf D_u\right)=\varepsilon^{|\mathbf D_u|}\,\vartheta^{m-|\mathbf D_u|}$. After calculating the expectation in $T$, we can determine the $T$ in this case as $T=\varepsilon^{|\mathbf D_u|}\,\vartheta^{m-|\mathbf D_u|}\Theta_{i}(\mathbf D_u,1)$. On the contrary, for the case where at least one of the $m-|\mathbf D_u|$ D2D pairs selects the overlay mode, we have $\beta\in[0,1]$ and $\mathbb P\left(\mathbf D_u\right)=\varepsilon^{|\mathbf D_u|}\left(\left(1-\varepsilon\right)^{m-|\mathbf D_u|}-\vartheta^{m-|\mathbf D_u|}\right)$. Calculating the expectation in $T$ yields $T=\varepsilon^{|\mathbf D_u|}\left(\left(1-\varepsilon\right)^{m-|\mathbf D_u|}-\vartheta^{m-|\mathbf D_u|}\right)\Theta_{i}(\mathbf D_u,\beta)$ in this case. Finally, substituting $T$ into (\ref{term_T}) yields (\ref{theorem_sop}).
 
Next, we proceed to prove the ASC in (\ref{theorem_asc}). According to the definition in (\ref{sc_def}), we have 
\begin{IEEEeqnarray}{rCl}
\mathbf C_s^{\,i} &=& \frac{1}{n}\mathbb E\left[\beta\left[\log\left(\frac{1+\frac{\mathrm{SNR}_{A_i,B}}{I_B+1}}{1+\frac{\mathrm{SNR}_{A_i,E}}{I_E+1}}\right)\right]^+\right]. 
\end{IEEEeqnarray}
As $\mathrm{SNR}_{A_i,B}$ and $\mathrm{SNR}_{A_i,E}$ are independent exponentially distributed random variables, we have
\begin{IEEEeqnarray}{rCl}
\mathbf C_s^{\,i} &=&\mathbb E\left[\frac{\beta}{n\ln2 }\!\left[\Psi\left(\frac{I_B\!+\!1}{\gamma_{A_i,B}}\!+\!\frac{I_B\!+\!1}{\gamma_{A_i,E}}\right)\!-\!\Psi\left(\frac{I_B\!+\!1}{\gamma_{A_i,B}}\right)\right]\right].
\end{IEEEeqnarray}
Based on the proof of SOP, we first apply the law of total expectation to change the above expectation into 
\begin{IEEEeqnarray}{rCl}\label{term_Q}
\mathbf C_s^{\,i} &=&\frac{1}{n \ln2 }\!\sum_{\mathbf D_u\in 2^{\mathcal D}}\mathbb P\left(\mathbf D_u\right)\mathbb E\Bigg[\beta\left[\Psi\left(\frac{I_B\!+\!1}{\gamma_{A_i,B}}\!+\!\frac{I_E\!+\!1}{\gamma_{A_i,E}}\right)\!-\!\Psi\left(\frac{I_B\!+\!1}{\gamma_{A_i,B}}\right)\right]\Big|\mathbf D_u\Bigg].
\end{IEEEeqnarray} 
We then divide the calculation of (\ref{term_Q}) into two cases of $\mathcal D_u=\mathcal D$ and $\mathcal D_u\neq \mathcal D$, and finally substitute the result of each case into (\ref{term_Q}) to yield (\ref{theorem_asc}). 

\section{Proof of Theorem \ref{Theorem_D2D}}\label{app_theorem_d2d}
We first prove the OP in (\ref{theorem_op}). It can be seen from (\ref{op_def}) that we need to derive $\mathbb P(R_{u}^{j,\,i}<r_t)$ and $\mathbb P(R_o^{\,j}<r_t)$ separately. First, we have 
\begin{IEEEeqnarray}{rCl}
\mathbb P(R_{u}^{j,\,i}<r_t)\!=\!\mathbb P\left(\frac{\mathrm{SNR}_{D_{j}^{t},D_j^{r}}}{\mathrm{SNR}_{A_i,D_{j}^{r}}\!+\!I_{D_j^{r}}^i\!+\!1}<2^{\frac{nr_t}{\beta }}\!-\!1\right)\!=\!1\!-\!\mathbb E\!\left[ \frac{e^{-\frac{\left(2^{\frac{nr_t}{\beta}}-1\right)\left(I_{D_j^{r}}^i+1\right)}{\gamma_{D_{j}^{t},D_j^{r}}}}}{\frac{\gamma_{A_i,D_{j}^{r}}}{\gamma_{D_{j}^{t},D_j^{r}}}\left(2^{\frac{nr_t}{\beta}}\!-\!1\right)\!+\!1}\right].
\end{IEEEeqnarray}
Following the idea of the proof in Theorem \ref{Theorem_CUE}, we obtain 
\begin{IEEEeqnarray}{rCl}\label{pod} 
\mathbb P(R_{u}^{j,\,i}<r_t)&=&1-\sum_{\mathbf D_u^j\in 2^{\mathcal D^{'}}} \varepsilon^{|\mathbf D_u^j|}\vartheta^{m-1-|\mathbf D_u^j|}\Omega_{i,j}(\mathbf D_u^j,1)\nonumber\\ 
&&\	\	-\sum_{\mathbf D_u^j\in 2^{\mathcal D^{'}}}\varepsilon^{|\mathbf D_u^j|}\Omega_{i,j}(\mathbf D_u^j,\beta)\left((1\!-\!\varepsilon)^{m\!-\!1\!-\!|\mathbf D_u^j|}\!-\!\vartheta^{m-1-|\mathbf D_u^j|}\right).
\end{IEEEeqnarray} 
Next, we calculate $\mathbb P(R_o^{\,j}<r_t)$, which can be given by
\begin{IEEEeqnarray}{rCl}\label{poc}
\mathbb P(R_o^{\,j}<r_t)&=&1\!-\!\sum_{l=1}^{m}\binom{m\!-\!1}{l\!-\!1}p^{m\!-\!l}(1\!-\!p)^{l-1}e^{-\frac{2^{\frac{lr_t}{(1-\beta)}}-1}{\gamma_{D_{j}^{t},D_j^{r}}}}.
\end{IEEEeqnarray}
Finally, substituting (\ref{pod}) and (\ref{poc}) into (\ref{op_def}) completes the proof of OP. 

We now proceed to prove the AC in (\ref{theorem_ac}). According to (\ref{ac_def}), we first calculate $\mathbb E[R_{u}^{j,\,i}]$ and then calculate $\mathbb E[R_o^{\,j}]$. Following the same idea in calculating the OP, $\mathbb E[R_{u}^{j,\,i}]$ can be given by
\begin{IEEEeqnarray}{rCl}\label{eqn_acd}
\mathbb E[R_{u}^{j,\,i}]&=&\sum_{\mathbf D_u^j\in 2^{\mathcal D^{'}}}\frac{\varepsilon^{|\mathbf D_u^j|+1}\Delta_{i,j}(\mathbf D_u^j)\left[\beta(1-\varepsilon)^{m-1-|\mathbf D_u^j|}+(1-\beta)\vartheta^{m-1-|\mathbf D_u^j|}\right]}{n\ln 2\left(\frac{\gamma_{A_i,D_{j}^{r}}}{\gamma_{D_{j}^{t},D_{j}^{r}}}-1\right)}
\end{IEEEeqnarray}
Next, $\mathbb E[R_o^{\,j}]$ can be given by 
\begin{IEEEeqnarray}{rCl}\label{eqn_acc}
\mathbb E[R_o^{\,j}]&=&-\frac{(1\!-\!\beta)\Psi\left(\frac{1}{\gamma_{D_j^{t},D_j^{r}}}\right)}{\ln 2}\sum_{l=1}^{m}\binom{m\!-\!1}{l\!-\!1}\frac{p^{m-l}(1\!-\!p)^{l\!-\!1}}{l }.
\end{IEEEeqnarray}
Finally, substituting (\ref{eqn_acd}) and (\ref{eqn_acc}) into (\ref{ac_def}) completes the proof.

\section{Proof of Lemma \ref{lemma_OP1_case}}\label{proof_lemma_OP1_case}
According to Lemma \ref{lemma_OP1}, the $\beta_1^*$ is given by $\beta_1^*=\max\left\{0,-\left(w_d\,\mu-w_c\nu \right)\frac{p}{1-p}+w_c\right\}$.
We can easily prove that $\nu\in(0,1)$, using the fact that $x\Psi(x)$ is an increasing function of $x$. 

Now, we continue to find the optimal $p_1^*$. First, consider case 1) in Lemma \ref{lemma_OP1_case}, where we have $0<\tilde{p}_1<\min\{1,w_c/(1-\nu)\}$ due to $\nu \in (0,1)$, where $\tilde{p}_1=\left(w_d/w_c\,\mu+(1-\nu)\right)^{-1}$. For the sub-region $p\in[0, \tilde{p}_1]$, the optimal $\beta$ is $\beta_1^*=-\left(w_d\,\mu-w_c\nu \right)\frac{p}{1-p}+w_c\in[0,w_c]$, which reduces $\mathcal C_s$ and $\mathbf R$ to  $\mathcal C_s=\frac{w_c \Lambda(\emptyset)}{\ln2}\left[\left(\mu-(1-\nu)\right)p+1\right]$ and $\mathbf R=-\frac{w_d \Psi\left(\frac{1}{\gamma_{D^{t},D^{r}}}\right)}{\ln2}\big[\left(\mu-(1-\nu)\right)p+1\big]$. As $\mu\geq 1-\nu$, $\mathcal U$ is an increasing function of $p$ in this sub-region. For the sub-region $p\in (\tilde{p}_1,1]$, the optimal $\beta$ is $\beta_1^*=0$, which reduces $\mathcal C_s$ and $\mathbf R$ to 
$\mathbf C_s = \frac{\Lambda(D)p}{\ln2 }$ and  $\mathbf R=\frac{\Psi\left(\frac{1}{\gamma_{D^{t},D^{r}}}\right)}{\ln 2}\big[\left(1-\nu\right)p-1\big]$.
The derivative of $\mathcal U$ in this case is $\frac{\partial{\mathcal U}}{\partial {p}}=\frac{w_c}{p}+\frac{w_d}{p-\frac{1}{1-\nu}}$. Thus, $\mathcal U$ increases with $p$ for $p\in (\tilde{p}_1, \frac{w_c}{1-\nu})$ and decreases with $p$ for $p \in(\frac{w_c}{1-\nu}, 1]$. Based on the above two sub-regions, the optimal $p$ in this case is thus $p_1^*=\min\{\frac{w_c}{1-\nu},1\}$ and the optimal $\beta$ is $\beta_1^*=0$.

Next, we consider case 2), where we have $\tilde{p}_1\geq 1$. Thus, for $p\in[0,1]$, the optimal $\beta$ is $\beta_1^*=\min\left\{-\left(w_d\,\mu-w_c\nu \right)\frac{p}{1-p}+w_c,1\right\}$. For the sub-region $p\in[0, \tilde{p}_2]$, where $\tilde{p}_2=(1-\mu+w_c/w_d\nu)^{-1}\in(\frac{w_d}{\nu},1]$, $\beta_1^*=-\left(w_d\,\mu-w_c\nu \right)\frac{p}{1-p}+w_c$. As $\mu \geq 1-\nu$, $\mathcal U$ is an increasing function of $p$ in this sub-region. For the sub-region $p\in(\tilde{p}_2,1]$, $\beta_1^*=1$, which reduces $\mathcal C_s$ and $\mathbf R$ to $\mathbf C_s = \frac{\Lambda(\emptyset)}{\ln2 }\left[\left(\mu-1\right)p+1\right]$ and  $\mathbf R=\frac{1}{\ln 2}\left(\frac{p\Delta(\emptyset)}{\frac{\gamma_{A,D^{r}}}{\gamma_{D^{t},D^{r}}}-1}\right)$. The derivative of $\mathcal U$ is $\frac{\partial{\mathcal U}}{\partial {p}}=\frac{w_c}{p+\frac{1}{\mu-1}}+\frac{w_d}{p}$. Thus, if $u<1$, we have $\tilde{p}_2<\frac{w_d}{1-\mu}$ and $\mathcal U$ is increasing in $p\in(\tilde{p}_2, \frac{w_d}{1-\mu})$ but decreasing in $p\in (\frac{w_d}{1-\mu}, 1]$. Combining the above two sub-regions, we can see that  if $u<1$, $(\beta_1^*,p_1^*)=(1, \min\{\frac{w_d}{1-\mu}, 1\})$. If $u\geq 1$, we have $\frac{\partial{\mathcal U}}{\partial {p}}>0$ and $\mathcal U$ is also increasing in $p\in(\tilde{p}_2,1]$. Thus, the optimal $p_1^*=1$ and the optimal $\beta_1^*$ can be any value in $[0,1]$, since $\mathcal U$ is independent of $\beta$ for $p=0$. 

Finally, for case 3) (i.e., $\mu+\nu<1$), we have $\frac{w_c}{1-\nu}<\tilde{p}_1$ if $w_c/\mu<w_d/\nu$ and $\frac{w_d}{1-\nu}<\tilde{p}_2$ if $w_c/\mu\geq w_d/\nu$. So, the objective function $\mathcal U$ always decreases with $p$ for $p\in [0,1]$. Thus, the optimal $p$ is $p_1^*=0$ and hence $\beta_1^*=w_c$ in this case. 

\bibliographystyle{IEEEtran}
\bibliography{MyReferences}

\begin{thebibliography}{10}
\providecommand{\url}[1]{#1}
\csname url@samestyle\endcsname
\providecommand{\newblock}{\relax}
\providecommand{\bibinfo}[2]{#2}
\providecommand{\BIBentrySTDinterwordspacing}{\spaceskip=0pt\relax}
\providecommand{\BIBentryALTinterwordstretchfactor}{4}
\providecommand{\BIBentryALTinterwordspacing}{\spaceskip=\fontdimen2\font plus
\BIBentryALTinterwordstretchfactor\fontdimen3\font minus
  \fontdimen4\font\relax}
\providecommand{\BIBforeignlanguage}[2]{{%
\expandafter\ifx\csname l@#1\endcsname\relax
\typeout{** WARNING: IEEEtran.bst: No hyphenation pattern has been}%
\typeout{** loaded for the language `#1'. Using the pattern for}%
\typeout{** the default language instead.}%
\else
\language=\csname l@#1\endcsname
\fi
#2}}
\providecommand{\BIBdecl}{\relax}
\BIBdecl

\bibitem{Boccardi2014}
F.~Boccardi, R.~W. Heath, A.~Lozano, T.~L. Marzetta, and P.~Popovski, ``Five
  disruptive technology directions for 5g,'' \emph{{IEEE} Commun. Mag.},
  vol.~52, no.~2, pp. 74--80, February 2014.

\bibitem{asadi2014survey}
A.~Asadi, Q.~Wang, and V.~Mancuso, ``A survey on device-to-device communication
  in cellular networks,'' \emph{{IEEE} Commun. Surveys Tuts.}, vol.~16, no.~4,
  pp. 1801--1819, 2014.

\bibitem{doppler2009d2d}
K.~Doppler, M.~Rinne, C.~Wijting, C.~B. Ribeiro, and K.~Hugl,
  ``Device-to-device communication as an underlay to lte-advanced networks,''
  \emph{{IEEE} Commun. Mag.}, vol.~47, no.~12, pp. 42--49, 2009.

\bibitem{fodor2012design}
G.~Fodor, E.~Dahlman, G.~Mildh, S.~Parkvall, N.~Reider, G.~Mikl{\'o}s, and
  Z.~Tur{\'a}nyi, ``Design aspects of network assisted device-to-device
  communications,'' \emph{{IEEE} Commun. Mag.}, vol.~50, no.~3, pp. 170--177,
  2012.

\bibitem{jo2015device}
M.~Jo, T.~Maksymyuk, B.~Strykhalyuk, and C.-H. Cho, ``Device-to-device-based
  heterogeneous radio access network architecture for mobile cloud computing,''
  \emph{{IEEE} Wireless Commun.}, vol.~22, no.~3, pp. 50--58, 2015.

\bibitem{camps2013device}
D.~Camps-Mur, A.~Garcia-Saavedra, and P.~Serrano, ``Device-to-device
  communications with wi-fi direct: overview and experimentation,''
  \emph{{IEEE} Wireless Commun.}, vol.~20, no.~3, pp. 96--104, 2013.

\bibitem{Mukherjee2014}
A.~Mukherjee, S.~Fakoorian, J.~Huang, and A.~Swindlehurst, ``Principles of
  physical layer security in multiuser wireless networks: A survey,''
  \emph{{IEEE} Commun. Surveys Tuts.}, vol.~16, no.~3, pp. 1550--1573, 2014.

\bibitem{NYang2015}
N.~Yang, L.~Wang, G.~Geraci, M.~Elkashlan, J.~Yuan, and M.~Di~Renzo,
  ``Safeguarding 5g wireless communication networks using physical layer
  security,'' \emph{{IEEE} Commun. Mag.}, vol.~53, no.~4, pp. 20--27, 2015.

\bibitem{Bassily2013SPM}
R.~Bassily, E.~Ekrem, X.~He, E.~Tekin, J.~Xie, M.~R. Bloch, S.~Ulukus, and
  A.~Yener, ``Cooperative security at the physical layer: A summary of recent
  advances,'' \emph{IEEE Signal Process. Mag.}, vol.~30, no.~5, pp. 16--28,
  Sept 2013.

\bibitem{zhang2015AHWSN}
Y.~Zhang, Y.~Shen, J.~Zhu, and X.~Jiang, ``Eavesdropper-tolerance capability in
  two-hopwireless networks via cooperative jamming.'' \emph{Adhoc \& Sensor
  Wireless Networks}, vol.~29, pp. 113--131, 2015.

\bibitem{Zhang2015TSC}
Y.~Zhang, Y.~Shen, H.~Wang, Y.~Zhang, and X.~Jiang, ``On secure wireless
  communications for service oriented computing,'' \emph{IEEE Transactions on
  Services Computing}, vol.~PP, no.~99, pp. 1--1, 2015.

\bibitem{Zhang2016TASE}
Y.~Zhang, Y.~Shen, H.~Wang, J.~Yong, and X.~Jiang, ``On secure wireless
  communications for iot under eavesdropper collusion,'' \emph{IEEE Trans.
  Autom. Sci. Eng}, vol.~13, no.~3, pp. 1281--1293, July 2016.

\bibitem{Tehrani2014IEEEComMag}
M.~N. Tehrani, M.~Uysal, and H.~Yanikomeroglu, ``Device-to-device communication
  in 5g cellular networks: challenges, solutions, and future directions,''
  \emph{{IEEE} Commun. Mag.}, vol.~52, no.~5, pp. 86--92, May 2014.

\bibitem{ChMaTCom2016}
C.~Ma, Y.~Li, H.~Yu, X.~Gan, X.~Wang, Y.~Ren, and J.~J. Xu, ``Cooperative
  spectrum sharing in d2d-enabled cellular networks,'' \emph{IEEE Trans.
  Commun.}, vol.~64, no.~10, pp. 4394--4408, Oct 2016.

\bibitem{Alihemmati2017TWC}
R.~AliHemmati, M.~Dong, B.~Liang, G.~Boudreau, and S.~Seyedmehdi,
  ``Multi-channel resource allocation towards ergodic rate maximization for
  underlay device-to-device communication,'' \emph{IEEE Trans. Wireless
  Commun.}, vol.~PP, no.~99, pp. 1--1, 2017.

\bibitem{Yue2013}
J.~Yue, C.~Ma, H.~Yu, and W.~Zhou, ``Secrecy-based access control for
  device-to-device communication underlaying cellular networks,'' \emph{{IEEE}
  Commun. Lett.}, vol.~17, no.~11, pp. 2068--2071, November 2013.

\bibitem{JWang2016}
J.~Wang, Q.~Tang, C.~Yang, R.~Schober, and J.~Li, ``Security enhancement via
  device-to-device communication in cellular networks,'' \emph{IEEE Signal
  Process. Lett.}, vol.~23, no.~11, pp. 1622--1626, Nov 2016.

\bibitem{HZhang2014ICC}
H.~Zhang, T.~Wang, L.~Song, and Z.~Han, ``Radio resource allocation for
  physical-layer security in d2d underlay communications,'' in \emph{Proc. IEEE
  ICC}, June 2014, pp. 2319--2324.

\bibitem{RZhang2016TWC}
R.~Zhang, X.~Cheng, and L.~Yang, ``Cooperation via spectrum sharing for
  physical layer security in device-to-device communications underlaying
  cellular networks,'' \emph{{IEEE} Trans. Wireless Commun.}, vol.~15, no.~8,
  pp. 5651--5663, Aug 2016.

\bibitem{LWang2015GLOBECOM}
L.~Wang, H.~Wu, M.~Peng, M.~Song, and G.~Stuber, ``Secrecy-oriented resource
  sharing for cellular device-to-device underlay,'' in \emph{Proc. IEEE
  GLOBECOM}, Dec 2015, pp. 1--5.

\bibitem{Zhang2017TVT}
K.~Zhang, M.~Peng, P.~Zhang, and X.~Li, ``Secrecy-optimized resource allocation
  for device-to-device communication underlaying heterogeneous networks,''
  \emph{IEEE Trans. Veh. Technol.}, vol.~66, no.~2, pp. 1822--1834, Feb 2017.

\bibitem{ChMa2015}
C.~Ma, J.~Liu, X.~Tian, H.~Yu, Y.~Cui, and X.~Wang, ``Interference exploitation
  in d2d-enabled cellular networks: A secrecy perspective,'' \emph{IEEE Trans.
  Commun.}, vol.~63, no.~1, pp. 229--242, 2015.

\bibitem{YLiu2016TOC}
Y.~Liu, L.~Wang, S.~A.~R. Zaidi, M.~Elkashlan, and T.~Q. Duong, ``Secure d2d
  communication in large-scale cognitive cellular networks: A wireless power
  transfer model,'' \emph{IEEE Trans. Commun.}, vol.~64, no.~1, pp. 329--342,
  Jan 2016.

\bibitem{Tolossa2017}
Y.~J. Tolossa, S.~Vuppala, G.~Kaddoum, and G.~Abreu, ``On the uplink secrecy
  capacity analysis in d2d-enabled cellular network,'' \emph{IEEE Syst. J.},
  vol.~PP, no.~99, pp. 1--11, 2017.

\bibitem{Sun2016TvT}
L.~Sun, Q.~Du, P.~Ren, and Y.~Wang, ``Two birds with one stone: Towards secure
  and interference-free d2d transmissions via constellation rotation,''
  \emph{IEEE Trans. Veh. Technol.}, vol.~65, no.~10, pp. 8767--8774, Oct 2016.

\bibitem{ye2014twc}
Q.~Ye, M.~Al-Shalash, C.~Caramanis, and J.~G. Andrews, ``Resource optimization
  in device-to-device cellular systems using time-frequency hopping,''
  \emph{{IEEE} Trans. Wireless Commun.}, vol.~13, no.~10, pp. 5467--5480, Oct
  2014.

\bibitem{lin2014TWC}
X.~Lin, J.~G. Andrews, and A.~Ghosh, ``Spectrum sharing for device-to-device
  communication in cellular networks,'' \emph{{IEEE} Trans. Wireless Commun.},
  vol.~13, no.~12, pp. 6727--6740, 2014.

\bibitem{Afzal2016WCNC}
A.~Afzal, S.~A.~R. Zaidi, D.~McLernon, and M.~Ghogho, ``On the analysis of
  device-to-device overlaid cellular networks in the uplink under 3gpp
  propagation model,'' in \emph{Proc. IEEE WCNC}, April 2016, pp. 1--6.

\bibitem{Chun2017TWC}
Y.~J. Chun, S.~L. Cotton, H.~S. Dhillon, A.~Ghrayeb, and M.~O. Hasna, ``A
  stochastic geometric analysis of device-to-device communications operating
  over generalized fading channels,'' \emph{{IEEE} Trans. Wireless Commun.},
  vol.~16, no.~7, pp. 4151--4165, July 2017.

\bibitem{zhu2015TWC}
K.~Zhu and E.~Hossain, ``Joint mode selection and spectrum partitioning for
  device-to-device communication: A dynamic stackelberg game,'' \emph{IEEE
  Trans. Wireless Commun.}, vol.~14, no.~3, pp. 1406--1420, 2015.

\bibitem{Huang2016TCOM}
Y.~Huang, A.~A. Nasir, S.~Durrani, and X.~Zhou, ``Mode selection, resource
  allocation, and power control for d2d-enabled two-tier cellular network,''
  \emph{IEEE Trans. Commun.}, vol.~64, no.~8, pp. 3534--3547, Aug 2016.

\bibitem{Feng2015TWC}
D.~Feng, G.~Yu, C.~Xiong, Y.~Yuan-Wu, G.~Y. Li, G.~Feng, and S.~Li, ``Mode
  switching for energy-efficient device-to-device communications in cellular
  networks,'' \emph{{IEEE} Trans. Wireless Commun.}, vol.~14, no.~12, pp.
  6993--7003, Dec 2015.

\bibitem{Lin2014CoMMag}
X.~Lin, J.~G. Andrews, A.~Ghosh, and R.~Ratasuk, ``An overview of 3gpp
  device-to-device proximity services,'' \emph{{IEEE} Commun. Mag.}, vol.~52,
  no.~4, pp. 40--48, 2014.

\bibitem{yu2011twc}
C.-H. Yu, K.~Doppler, C.~B. Ribeiro, and O.~Tirkkonen, ``Resource sharing
  optimization for device-to-device communication underlaying cellular
  networks,'' \emph{{IEEE} Trans. Wireless Commun.}, vol.~10, no.~8, pp.
  2752--2763, 2011.

\bibitem{Cheng2016JSAC}
W.~Cheng, X.~Zhang, and H.~Zhang, ``Optimal power allocation with statistical
  qos provisioning for d2d and cellular communications over underlaying
  wireless networks,'' \emph{IEEE J. Sel. Areas Commun.}, vol.~34, no.~1, pp.
  151--162, 2016.

\bibitem{Mdval5}
\BIBentryALTinterwordspacing
C++ simulator for mode selection and spectrum partition in d2d communications
  underlaying cellular networks with phy security consideration. [Online].
  Available: \url{http://mdlval.blogspot.jp/}
\BIBentrySTDinterwordspacing

\end{thebibliography}


%

%
%

\ifCLASSOPTIONcaptionsoff
  \newpage
\fi

\end{document}